\documentclass[twocolumn, trackchanges]{aastex62}
\usepackage{enumitem}
\usepackage{color}
\usepackage{amsmath}
\usepackage{newtxtext,newtxmath}
\usepackage{xcolor}
\usepackage{graphicx}
\usepackage{footnote}
\usepackage{amssymb}
\usepackage{multirow}
\def\fsj151857{{Swift J151857.0$-$572147}}
\def\sj151857{{J151857}}

\def\ngso{{\it Neil Gehrels Swift Observatory}}
\def\swift{{\it Swift}}
\def\nustar{{\it NuSTAR}}

\def\nicer{{\it NICER}}
\def\astrosat{{\it AstroSat}}
\def\meerkat{{\it MeerKAT}}
\def\insight{{\it Insight}}

\newcommand{\erg}{erg cm$^{-2}$ s$^{-1}$} 
\newcommand{\lum}{erg s$^{-1}$} 
\newcommand{\lxp}{\emph{AstroSat}/LAXPC}
\newcommand{\sxt}{\emph{AstroSat}/SXT}
\newcommand{\czti}{\emph{AstroSat}/CZTI}

\def\src{{Swift J151857.0$-$572147}}

\usepackage{hyperref}
\shorttitle{Multiwavelength observation of \src~}
\shortauthors{Mandal et al.}

\begin{document}
\title{\large \bf Multiwavelength View of Black Hole X-ray Binary \src\ 
}

\correspondingauthor{Manoj Mandal}
\email{manojmandal213@gmail.com}

\author[0000-0002-1894-9084]{Manoj~Mandal}
\affil{Astronomy and Astrophysics Division, Physical Research Laboratory, Navrangpura, Ahmedabad - 380009, Gujarat, India}

\author[0000-0003-2865-4666]{Sachindra Naik} 
\affil{Astronomy and Astrophysics Division, Physical Research Laboratory, Navrangpura, Ahmedabad - 380009, Gujarat, India}

\author[0000-0002-6794-7405]{Souvik Manik}
 \affil{Department of Pure and Applied Sciences, Midnapore City College, Kuturia, Bhadutala, West Bengal, 721129, India}
 \affil{Space, Planetary \& Astronomical Sciences \& Engineering (SPASE), IIT Kanpur, 208016, India}
 
 \author[0000-0003-2325-8509]{Sabyasachi Pal}
 \affil{Department of Pure and Applied Sciences, Midnapore City College, Kuturia, Bhadutala, West Bengal, 721129, India}
 
\begin{abstract}
We investigate multiwavelength timing and spectral properties of the newly discovered black hole X-ray binary \src{} during its 2024 outburst using observations from \astrosat, \nustar, \nicer, \swift, and \meerkat. The \astrosat{}/LAXPC20 power density spectrum reveals a $\sim$8 Hz low-frequency QPO with a quality factor of $\sim2.8$. The broadband X-ray spectrum of the source exhibits prominent reflection features, which are modeled to constrain the geometry and physical properties of the accretion disk. The spectral analysis indicates that the source was in the soft/intermediate spectral state, characterized by a steep power law with a photon index of $\Gamma \approx 2.6$ and a relatively high disk temperature of $\sim0.9$~keV. The reflection modeling suggests a disk inclination of $23-31^{\circ}$ and a highly ionized accretion disk with $\log\xi \sim 3.3$--4.0. The \meerkat{} monitoring revealed a strong radio variability that closely tracked the hard X-ray evolution, including a bright radio flare during the hard-to-soft state transition, supporting close coupling between the accretion flow and jet activity. Using the quasi-simultaneous \astrosat{}, \swift/XRT and \meerkat{} observations, we investigated the radio/X-ray correlation and found that \src{} follows the established $L_{\rm X}$--$L_{\rm R}$ relation for Galactic black hole X-ray binaries. Its location in the $L_{\rm X}$--$L_{\rm R}$ plane is consistent with that of Galactic black hole X-ray binaries, further supporting its black hole nature.
\end{abstract}

\section{Introduction}
\label{sec:intro}
The X-ray sky exhibits variability over a wide range of timescales, with transient phenomena providing valuable opportunities to study rapidly changing accretion processes. Accreting compact objects in binary systems display diverse spectral and timing behaviour, including state transitions, quasi-periodic oscillations, relativistic reflection, time lags, and transient jets \citep{rem06}. These systems therefore provide powerful laboratories for investigating accretion and ejection processes in the strong-gravity regime. Black hole X-ray binaries (BHXRBs) consist of stellar-mass black holes accreting matter from the companion stars through the accretion disk. Depending on the mass of the donor star, BHXRBs are broadly classified as low-mass and high-mass binary systems \citep{rem06}. The majority of transient BHXRBs belong to the low-mass binary category, where the companion star typically has a mass lower than that of the Sun. During an X-ray outburst, the luminosity of these systems increases by several orders of magnitude and can be detected across the electromagnetic spectrum \citep{rem06}. The evolution of an outburst is commonly traced using the hardness--intensity diagram (HID), which reveals a sequence of accretion states characterized by distinct spectral and timing properties \citep{Homan2005, rem06}.

At the onset of an outburst, the BHXRBs are generally observed in the hard state (HS), where the X-ray emission is dominated by the non-thermal component arising from Comptonization in a hot corona and, possibly, contributions from the jet. The thermal emission from the accretion disk is relatively weak in this spectral state \citep{Belloni2005}. As the mass accretion rate increases, the source evolves through the hard-intermediate state (HIMS) and soft-intermediate state (SIMS). The source subsequently enters the soft state (SS), where the thermal emission from the accretion disk dominates the spectrum, and the disk is expected to extend close to the innermost stable circular orbit (ISCO) \citep{rem06}. In this spectral state, the contribution from the non-thermal component is typically less than \(\sim25\%\) of the total X-ray emission. During the decline phase of the outburst, the source retraces the sequence of spectral states in reverse order, returning from the soft state to the hard state. The transition from the hard to soft state during the rise and from the soft to hard state during the decay follows a hysteresis pattern in the HID, producing the characteristic q-shaped track associated with a complete outburst \citep{Homan2001,Belloni2005}. 

The evolution of the spectral states is accompanied by significant changes in the X-ray timing properties, providing valuable insights into the geometry and dynamics of the accretion flow. Among the various timing features, low-frequency quasi-periodic oscillations (QPOs) are particularly important, as their properties systematically evolve with the spectral state of the source \citep{rem06}. Based on their timing characteristics, such as centroid frequency, quality factor ($Q$), fractional rms amplitude, and underlying broadband noise, the low-frequency QPOs are generally classified into three categories: type-A, type-B, and type-C \citep{Casella2005}. During the hard state (HS) and hard-intermediate state (HIMS), the X-ray emission is dominated by the hard spectral component, and strong type-C QPOs are commonly observed with evolving timing properties. In contrast, the soft-intermediate state (SIMS) and soft state (SS) are dominated by thermal disk emission. Type-A and type-B QPOs may appear intermittently during the SIMS, whereas QPOs are generally absent in the SS, although the rare occurrence of type-A QPOs has also been reported in this state. Type-A QPOs are characterized by a weak, broad, and nearly flat-topped feature at frequencies of $\sim$6--8 Hz, with fractional RMS amplitudes of only a few per cent \citep{Homan2001}. Type-C QPOs are strong and narrow features typically observed over a frequency range of $\sim$0.1--15 Hz, with high quality factors ($Q \sim 7$--12) and large fractional rms amplitudes ($\sim$3--16\%). Type-B QPOs generally occur near 5--6 Hz, with $Q \gtrsim 6$ and fractional rms amplitudes of $\sim$2--4\%. In contrast, type-A QPOs are rare, broad features typically with both the quality factor and fractional rms amplitude $\lesssim 3$ \citep{Casella2005}. In this case, harmonics of the QPO are generally absent, likely because of the broad nature of the QPO.

Timing evolution of BHXRBs is closely coupled to changes in their X-ray spectral properties, reflecting variations in the geometry and physical conditions of the accretion flow. The X-ray spectra of BHXRBs are generally described by two primary components: a multicolor thermal blackbody originating from the accretion disk and a hard non-thermal power-law component produced through inverse Compton scattering in the corona. Spectral parameters provide important diagnostics of the accretion state. The hard state is characterized by a flatter power-law spectrum with \(\Gamma \sim 1.5-1.8\) and a cooler accretion disk. In the soft state, the X-ray spectrum is characterized by a steep photon index ($\Gamma \gtrsim 2.5$) and a relatively hot accretion disk with an inner disk temperature of $\sim1$~keV \citep{rem06,sha73, Jana2020, Saha2023, Mandal2024}. The emission is dominated by a strong thermal disk blackbody component below $\sim10$~keV, accompanied by a weak high-energy tail that contributes only $\sim25\%$ of the total bolometric luminosity \citep{Done2007,You2021}. These systems emit radiation across the electromagnetic spectrum, from radio to X-rays. Multiwavelength observations play a crucial role in understanding the emission mechanisms operating in BHXRBs. In particular, simultaneous radio and X-ray observations have established a strong connection between accretion processes and jet production, providing key insights into the coupling between the accretion flow and relativistic outflows \citep{Corbel2003}. 

The X-ray transient \fsj151857\ was discovered by the \ngso\ \citep[\swift;][]{ken24}. Initially identified as a possible gamma-ray burst, it was later classified as a new X-ray transient exhibiting an X-ray flux of $\sim500$~mCrab, a high hydrogen column density ($\sim5\times10^{22}$~cm$^{-2}$), and a hard X-ray spectrum ($\Gamma\sim1.8$). Radio observations with the \meerkat{} telescope detected the source with high significance, revealing a flux density and spectral index consistent with an X-ray binary in the hard state, suggesting either a black hole or a radio-bright neutron-star system \citep{cow24}. Subsequent observations with the Australia Telescope Compact Array \citep[{\it ATCA};][]{car24} detected exceptionally bright radio flares, characteristic of Galactic black hole X-ray binaries launching relativistic ejecta during the hard-to-soft state transition. Further \swift/XRT observations of \src{} confirmed a likely transition of the source to a soft state on 2024 March 10 \citep{DelSanto2024}. Owing to strong extinction, no optical counterpart has been identified \citep{pay24}; however, a near-infrared counterpart was reported by \cite{bag24} from H-band observations obtained with the 60~cm Robotic Eye Mount (REM) telescope at La~Silla, Chile. Additionally, \insight{}-HXMT observations revealed type-C QPOs from the source during its intermediate state \citep{Chatterjee2025}. A detailed broadband spectral analysis using \insight{}-HXMT, \nicer, and \nustar{} observations was subsequently carried out to constrain the accretion geometry and estimate the black hole spin and mass \citep{Peng2024}.

In this paper, we investigate the multiwavelength timing and spectral properties of \src{} using observations from \astrosat, \nicer, \swift, \nustar, and \meerkat. We report the detection of an $\sim$8 Hz quasi-periodic oscillation (QPO) and present a detailed timing analysis using \astrosat. The broadband X-ray spectrum exhibits prominent reflection features, which are modeled with relativistic reflection models to constrain the geometry of the inner accretion flow. We further incorporate radio observations to investigate the accretion--jet connection and examine the source's location in the radio/X-ray luminosity plane. The paper is organized as follows. Section~\ref{sec:obs} describes the observations and data reduction procedures, Section~\ref{sec:results} presents the results, and Section~\ref{sec:discon} summarizes the discussion and conclusions.

\begin{deluxetable*}{cccccccc}
\tabletypesize{\footnotesize}
\tablecaption{Observation log of the radio and X-ray observations of \src{} during the 2024 outburst. The quoted radio flux densities are measured at 1.28 GHz L-band of \meerkat{}.}
\label{tab:obs}
\tablecolumns{7}
\tablewidth{0pt}
\tablehead{
\colhead{Observatory} & \colhead{Obs. ID} & \colhead{Obs. Date} & \colhead{Exposure (ks)} & Flux density (mJy) \vspace{0.3cm}
} 
\startdata
\astrosat\     & T05\_088T01\_9000006126  &  2024-03-16  (MJD 60385.87) &  83.2  & ------         \\
\nustar\ & 91001311004 &  2024-03-18 (MJD 60387.65)  &9.2    & ------\\
\nicer\     & 7204220111  &  2024-03-18 (MJD 60387.55) & 4.2  &  ------                         \\
 \swift{}/XRT & 01218604000 &  2024-03-04 (MJD 60373.36)  &1.7    & ------\\
\swift{}/XRT & 00016560001 &  2024-03-10 (MJD 60379.31)  &1.0    & ------\\
\swift{}/XRT & 00016560014 &  2024-04-07 (MJD 60407.71)  &0.7    & 
------                        \\
\hline
\meerkat\     & 1709516964  &  2024-03-04  (MJD 60373.08) &  2.79 & $10.33\pm0.05$         \\
\meerkat\     & 1709952327  &  2024-03-09  (MJD 60378.11) &  7.21 &   $265.02\pm0.89$       \\
\meerkat\     & 1710035142  &  2024-03-10  (MJD 60379.07) &  2.80  & $210.10\pm0.12$        \\
\meerkat\     & 1710551895  &  2024-03-16  (MJD 60385.06) &  9.70  &  $31.28\pm0.12$       \\
\meerkat\     & 1711241475  &  2024-03-24  (MJD 60393.04) &  7.41  &  $12.70\pm0.05$       \\
\meerkat\     & 1712527817  &  2024-04-07  (MJD 60407.93) &  6.97  &  $21.46\pm0.09$        \\
\hline
\enddata
\end{deluxetable*}

\section{Observations and Data Reduction}
\label{sec:obs}
In this section, we summarize observations of \src~ included in this work (Table~\ref{tab:obs}) and describe the observational strategy, data reduction, and analysis procedures adopted for each instrument and observatory.

\subsection{\astrosat}\label{sec:ix}
\astrosat\ is the first dedicated multiwavelength astronomy satellite of India, which was launched in 2015. Onboard \astrosat, there are five main payloads: the Soft X-ray Telescope (SXT), the Large Area X-ray Proportional Counters (LAXPCs), the Cadmium-Zinc-Telluride Imager (CZTI),  the Ultra-Violet Imaging Telescope (UVIT), and the Scanning Sky Monitor (SSM). The details on \astrosat~ are given in \citet{Agrawal2006, Singh2014}. Following the discovery of the new X-ray transient \src, we proposed an \astrosat~ Target of Opportunity (ToO) observation (ObsID T05\_088T01\_9000006126) for an exposure of $\sim$80 ks to understand the broadband nature of the source. The ToO observation of \src{} was carried out on 2024 March 16-18.
 \subsubsection{\sxt{}}
The SXT is a focusing X-ray telescope with a CCD at the focal plane \citep{Singh2016, Singh2017}. It is sensitive in the 0.3-8 keV range. \src~ was observed in the Photon Counting (PC) mode for an exposure of $\sim$42 ks. The \sxt{} event merger tool (Julia Code) was used to merge the level~2 cleaned files from individual orbits. Using the \textsc{xselect} package in \textsc{heasoft}, the combined event file was used to extract the image, light curve, and spectrum of the source. The source count rate ($\sim67.5$ counts/s) was significantly higher than the pileup threshold of 40 counts/s. The data are corrected for the photon pile-up by considering an annular region with inner and outer radii of 4 and 12 arcmin, respectively, centered at the source position. The spectral redistribution matrix file (sxt\_pc\_mat\_g0to12.rmf), provided by the \sxt{} team\footnote{\url{http://www.tifr.res.in/~astrosat\_sxt/dataanalysis.html}}, and the blank sky \sxt{} spectrum as background (SkyBkg\_sxt\_LE0p35\_R16p0\_v05\_Gd0to12.pha) are utilized for spectral analysis. Using the {\tt sxtARFModule} tool, we generated the off-axis auxiliary response files (ARF) based on the provided on-axis ARF (sxt\_pc\_excl00\_v04\_20190608.arf). For spectral modeling, we used data in the 0.9-7 keV range and added a systematic error of 2\% to the spectrum. While fitting, data in the range of 1.5--3.0~keV are excluded due to known instrumental features \citep{Sharma2024,Chhotaray2023, Jana2022}. Since our broadband analysis includes \nicer\ data, excluding these channels from \sxt~ data does not affect the spectral fitting.
\subsubsection{\lxp{}}
One of the primary instruments on board \astrosat~ is the Large Area X-ray Proportional Counter (LAXPC). It consists of three identical co-aligned proportional counters (LAXPC10, LAXPC20, and LAXPC30) that operate in the 3-80 keV energy range, with a total effective area of 8000 cm$^2$ at 15 keV. The timing resolution of the instruments is 10 $\mu$s, whereas the spectral resolution is 12\% at 22 keV \citep[for details see][]{Yadav2016, Antia2017}. We used data from the LAXPC20 detector in our analysis, as LAXPC10 was running at low gain, and LAXPC30 had been turned off since 8 March 2018 due to irregular gain changes. The source was observed with the LAXPC for an effective exposure of $\sim$38 ks. We used standard data analysis tools {\tt LAXPCsoftware} (laxpcsoft; version 2020 August 4) to extract source light curves and spectra. The background estimation was performed using the blank sky observations closest to the \src~ observation. The response file is created as described in Antia et al. (2017). As the nearby source Cir~X-1 was in the field of view of \lxp{}, there is a possibility of contamination of \src~ data in the soft X-ray range. For this reason, we used data in the 5-25 keV range in our spectral fitting. Data beyond 25 keV are not considered, as it is background-dominated. A systematic error of 2\% is added to the \lxp{} spectra, as recommended by the instrument team.
\subsubsection{\czti{}}
The Cadmium Zinc Telluride Imager (CZTI) is a hard X-ray imaging instrument that uses a coded-mask aperture to subtract background radiation. It is sensitive in the 20–200 keV energy range \citep{Bha17}. Sixteen pixelated detector modules are distributed across its four quadrants. With a field of view of 4.6$^{\circ}$ $\times$ 4.6$^{\circ}$, \czti{} provides an angular resolution of 8 arcmin. \src was observed with \czti{} for an exposure of $\sim$83 ks. The \czti{} data analysis pipeline version~3.0 was used to analyze the \czti{} data. We created level-2 cleaned event files from the level-1 data using the \texttt{cztpipeline} tool. The \texttt{cztbindata} task was used to extract the source spectra. \texttt{cztaddspec} was used to add the spectra (20-90 keV range) from all four \czti{} quadrants. 
 \begin{figure}
\centering{
\includegraphics[width=8.5 cm]{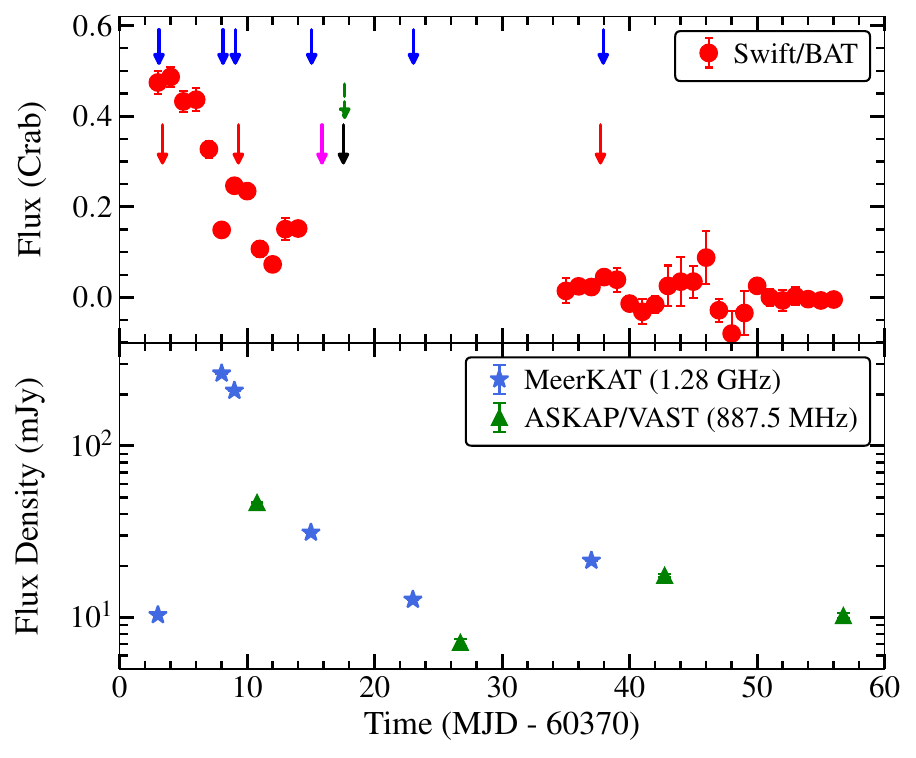}
\caption{Top panel: Hard X-ray light curve of \src{} obtained with \swift/BAT (15--50 keV) during its 2024 outburst. The \astrosat, \nicer{}, \swift/XRT, \nustar{}, and \meerkat{} observation epochs are marked with magenta, black, red, green, and blue arrows, respectively. Bottom panel: Estimated radio flux densities from the \meerkat\ observations at 1.28 GHz are shown during the same X-ray outburst. The radio flux density measurements from {\it ASKAP}/VAST are also shown with green triangles and are taken from \citet{Anumarlapudi24}. As these measurements were obtained predominantly during the declining phases of the outburst, when the ejecta is expected to be less self-absorbed, we assume a flat spectral index between the VAST and \meerkat\ bands and apply no frequency scaling between the two datasets.}
\label{fig:BAT}}
\end{figure}
 \begin{figure*}
\centering{
\includegraphics[width=6.5 cm, angle=270, trim=0cm 2.25cm 0cm 0cm, clip]{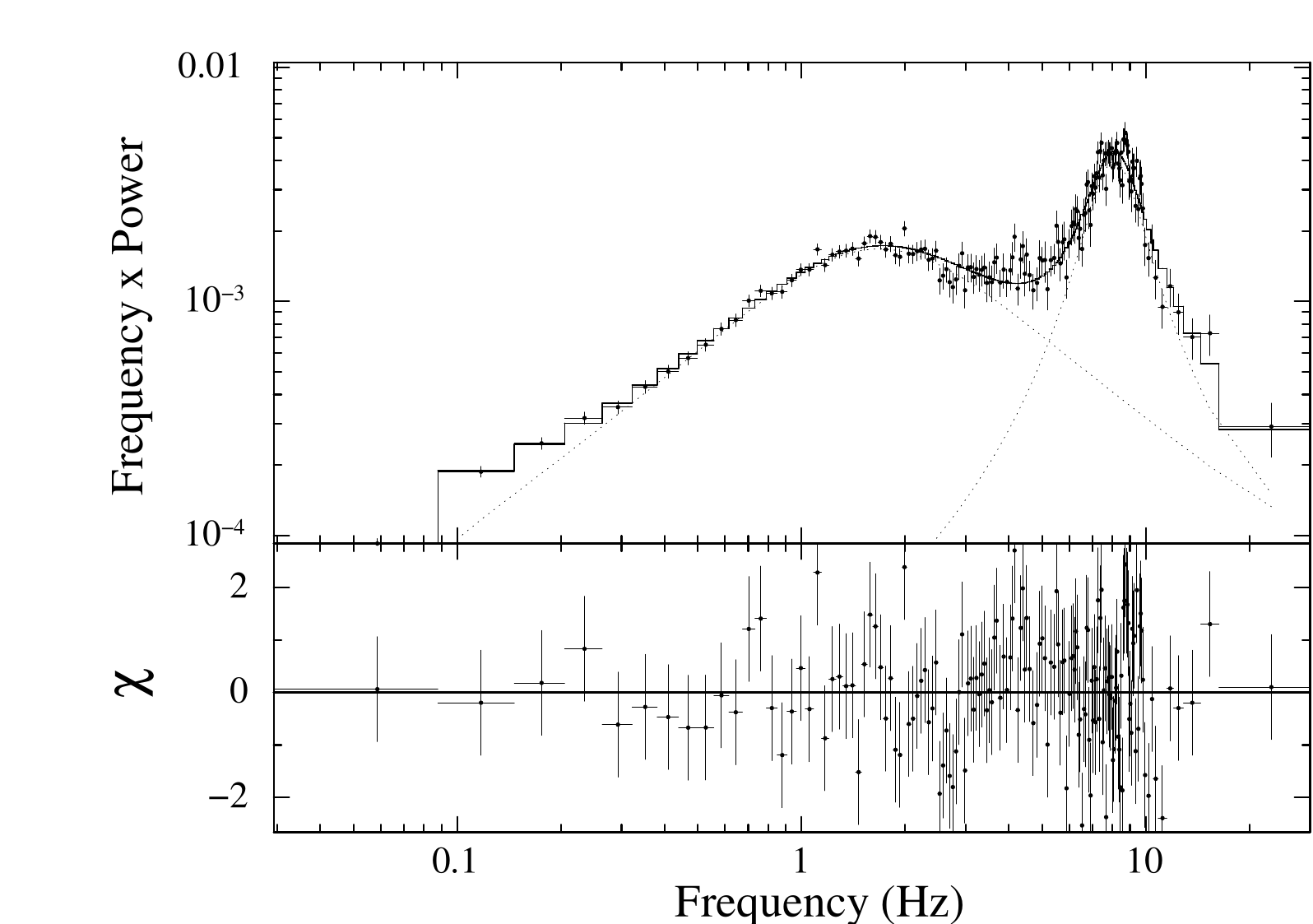}
\includegraphics[width=6.5 cm, angle=270]{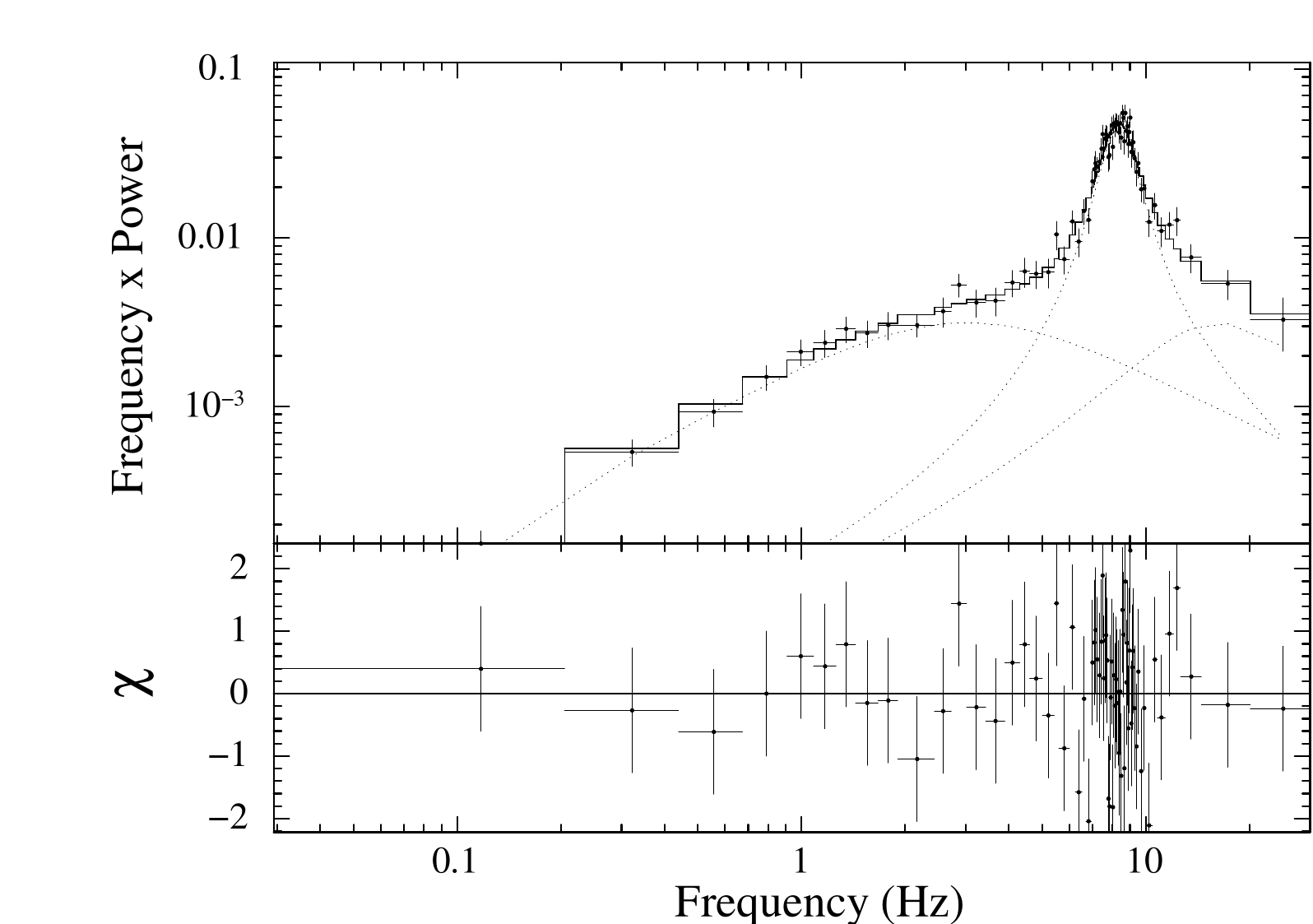}
\caption{{Power density spectra (PDS) of \lxp{}20 in the 3--10 keV (left panel) and 10--25 keV (right panel) energy bands. The black data points represent the observed PDS, while the black dotted lines show the best-fit models. The lower panels display the fit residuals.} 
}
\label{fig:PDS}}
\end{figure*}
\subsection{\nicer}
\label{sec:ni}
The Neutron Star Interior Composition Explorer (\nicer; \citealt{Gendreau2016}) is an X-ray timing mission operating in the 0.2--12~keV energy band, providing a large effective area and excellent time resolution for studies of compact objects. \nicer{} monitored \src{} regularly during its 2024 outburst, with observations beginning on 2024 March 18. \nicer~ observations were conducted using 50 active focal-plane modules (FPMs), with FPMs 14 and 34 excluded from the \nicer\ 52-module array. The \nicer~ observation (Obs. ID-7204220111 with an exposure of $\sim$4.2 ks) closest in time to the \astrosat~ and \nustar~ observations was selected for the broadband spectral analysis. The details of the observation are summarized in Table~\ref{tab:obs}. We therefore performed broadband spectral analysis of these quasi-simultaneous datasets over the 0.9--90~keV energy range. The raw data are processed using the tool {\tt NICERDAS} in {\tt HEASoft} with the {\tt CALDB} version xti20240206. The {\tt nicerl2} tool is utilized to filter the raw event file. The {\tt XSELECT} package is used to generate light curves and spectra from the clean event file. The corresponding background spectrum is generated using the {\tt nibackgen3C50} tool \citep{Re22}. The response and ancillary response files are generated using {\tt nicerrmf} and {\tt nicerarf}. The spectrum is grouped considering optimal binning \citep{2016A&A...587A.151K}, with 30 counts per grouped bin. For spectral modeling, a systematic error of 1.5\% is added to the \nicer~ spectra, as recommended by the instrument team\footnote{\url{https://heasarc.gsfc.nasa.gov/docs/nicer/analysis_threads/cal-recommend/}}. 
\subsection{\swift{}/XRT}
\label{sec:xrt}
\src~ was observed with the \textit{Swift}/X-ray Telescope (XRT; \citet{Gehrels2004}) during its 2024 outburst. In this work, we used three epochs of \textit{Swift}/XRT observations that are simultaneous/quasi-simultaneous with the radio observations. The details of the observations are given in Table~\ref{tab:obs}. All observations were performed in Windowed Timing (WT) mode. The data were processed using \texttt{HEASOFT} with CALDB version 20250609 and reduced with \texttt{XRTPIPELINE} (v0.13.7). Source light curves and spectra were extracted from a $40\times20$ pixel rectangular region, while background products were obtained from a nearby source-free region of similar size using \texttt{XSELECT} (v2.5c). Exposure maps and ARFs were generated with \texttt{XRTEXPOMAP} and \texttt{XRTMKARF}, respectively, using the latest CALDB response files. The spectra were grouped to a minimum of 25 counts per bin using \texttt{grppha}, ensuring sufficient counts per bin for $\chi^2$ statistics.
\subsection{NuSTAR}
\label{sec:nu}
The Nuclear Spectroscopic Telescope Array (\nustar{}) observatory comprises two co-aligned focal plane modules, FPMA and FPMB, providing sensitive imaging and spectroscopy over the 3--79~keV energy range \citep{Ha13}. \nustar\ observed \src\ multiple times during the 2024 outburst. In this work, we selected the observation closest in time to the quasi-simultaneous \nicer\ and \astrosat\ observations to facilitate a broadband spectral and timing analysis. We used a \nustar\ observation of \src\ on 2024 March 18 with a total exposure of $\sim9$~ks (see Table~\ref{tab:obs}). The data were processed using the {\tt NUSTARDAS} software package included in {\tt HEASoft} v6.36, together with the latest calibration database ({\tt CALDB} v20250428). Cleaned event files were generated with {\tt NUPIPELINE}. Source events were extracted from a circular region of radius 120~arcsec centered on the target, while background events were obtained from a nearby source-free region. Finally, light curves and spectra for both FPMA and FPMB were generated using the {\tt NUPRODUCTS} task. The spectra were grouped using {\tt ftgrouppha} with optimal binning and a minimum of 30 counts per spectral bin.
 \begin{figure}
\centering{
\includegraphics[width=9.0 cm, trim=0.75cm 0cm 0cm 0cm, clip]{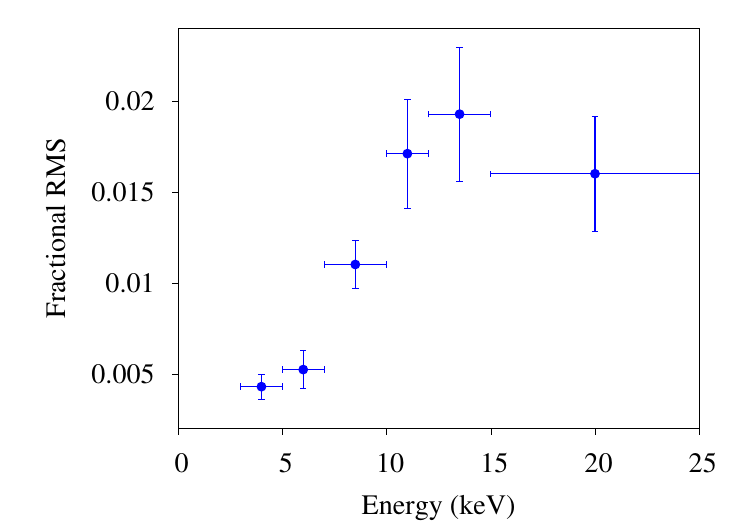}
\caption{Energy dependence of the intrinsic fractional rms amplitude at the $\sim$8~Hz QPO frequency, computed using the \texttt{laxpc\_find\_freqlag} task. The intrinsic fractional rms was estimated from the deadtime- and background-corrected Fourier power at the selected QPO frequency ($\sim$8.1~Hz) and is shown as a function of energy in the 3--25~keV band.}
\label{fig:frac_rms}}
\end{figure}
\subsection{Radio observations with \meerkat}
Among the six epochs of \meerkat~ radio observations of \src, one observation is quasi-simultaneous with the \astrosat~ X-ray observations. In addition, there are three \swift/XRT observations that are quasi-simultaneous with the \meerkat~ epochs, used in the present work. Details of the observations are summarized in Table~\ref{tab:obs}. The \meerkat~ observations were conducted at a central frequency of 1.28 GHz (L-band), spanning 856--1712 MHz across 32768 channels with 8~s integrations, using 64 \textit{MeerKAT} antennas. To reduce data volume and computational complexity, we averaged the visibility data across 32768 to 8192 channels. We carried out the initial flagging in CASA (\citealt{2007ASPC..376..127M}; CASA Team 2022), removing the band-edge channels, known bad channels, and visibilities with non-physical amplitudes. We then performed a second round of flagging on the calibrator fields to identify and remove the remaining radio frequency interference (RFI) in the time--frequency domain. 

Next, we derived the flux density, bandpass, and delay calibration solutions from the primary calibrator and determined the complex gain solutions from the secondary calibrator. After applying the calibration solutions, we split the calibrated visibilities into a separate target measurement set. We subsequently removed the residual RFI from the target data using \texttt{TRICOLOUR} \citep{2022ASPC..532..541H}, which is optimized for wideband \textit{MeerKAT} observations. We produced an initial continuum image with \texttt{WSClean} (\citealt{2014MNRAS.444..606O,2017MNRAS.471..301O}) using multi-frequency synthesis and multi-scale CLEAN, and we used this image to generate a deconvolution mask. We then performed phase-only self-calibration with \texttt{CubiCal} \citep{Kenyon2018} before producing the final continuum image. The integrated flux density and its associated uncertainty of \src\ from each image were estimated by fitting a two-dimensional Gaussian using the CASA task \texttt{IMFIT}.
\begin{table}
	\centering
	\caption{Best-fit Lorentzian parameters for the QPO detected in the \astrosat\ \lxp{}20 power spectra in the 3--25, 3--10, and 10--25 keV bands. Uncertainties are quoted at the 90\% confidence level for a single parameter.}
	\resizebox{\linewidth}{!}{
\begin{tabular}{lccc}
\hline
 QPO Parameters &   3-25 keV  & 3-10 keV & 10-25 keV \\
\hline
 Freq., $\nu_0$ (Hz) & $8.07 \pm 0.04$ & $7.95 \pm 0.07$ & $8.26 \pm 0.06$ \\[1ex]
FWHM, $\Delta$ (Hz) & $2.96^{+0.14}_{-0.13}$ & $3.22^{+0.23}_{-0.22}$ & $2.25^{+0.14}_{-0.33}$ \\[1ex]
Integrated QPO rms (\%) &  $5.68^{+0.10}_{-0.09}$ & $4.84\pm0.13$ & $13.5^{+0.5}_{-1.1}$\\[1ex]
Quality Factor, $Q$ & $2.73\pm0.14$ &  $2.47^{+0.20}_{-0.19}$ &  $3.68^{+0.66}_{-0.25}$ \\[1ex]   
\hline
$\chi^2/{\rm dof}$ &  242/173 & 181/152 & 65/64\\
\hline
\end{tabular}
}
\label{tab:pds}
\end{table}
\begin{figure}
\centering{
\includegraphics[width=0.45\columnwidth, angle=-90, trim=0cm 1.15cm 0cm 0cm, clip]{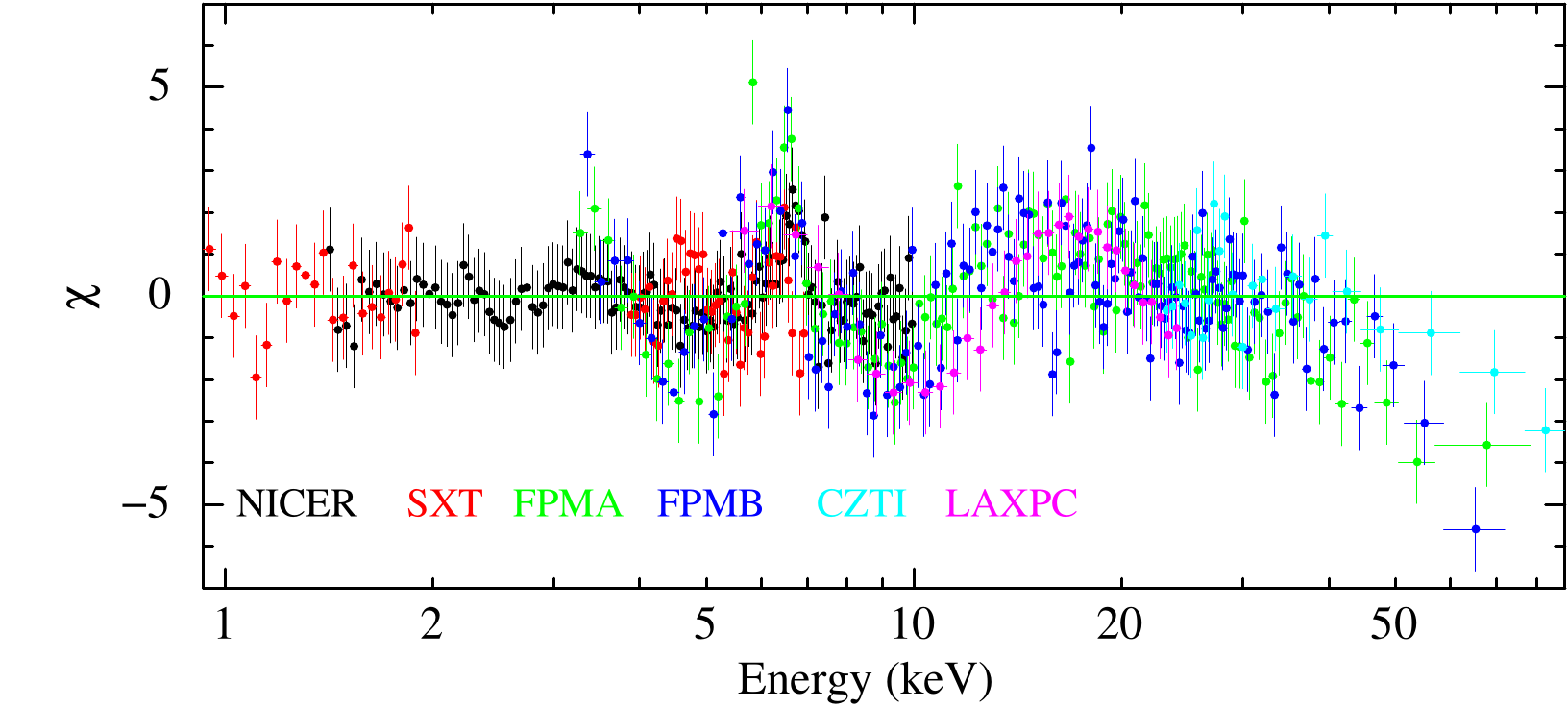}
\caption{The residuals of the broadband spectra fitted with \texttt{tbabs $\times$ (powerlaw + diskbb)} model reveal the presence of clear reflection features, including an Fe~K$\alpha$ emission line at $\sim$6.4~keV and a broad Compton hump at around $\sim$20~keV. The presence of these residuals indicates that a reflection component is required to adequately describe the source spectrum.}
\label{fig:residual_plot}}
\end{figure}
\begin{figure}
\centering
 \includegraphics[width=0.33\textwidth, angle=270, trim=0cm 1.3cm 0cm 0cm, clip]{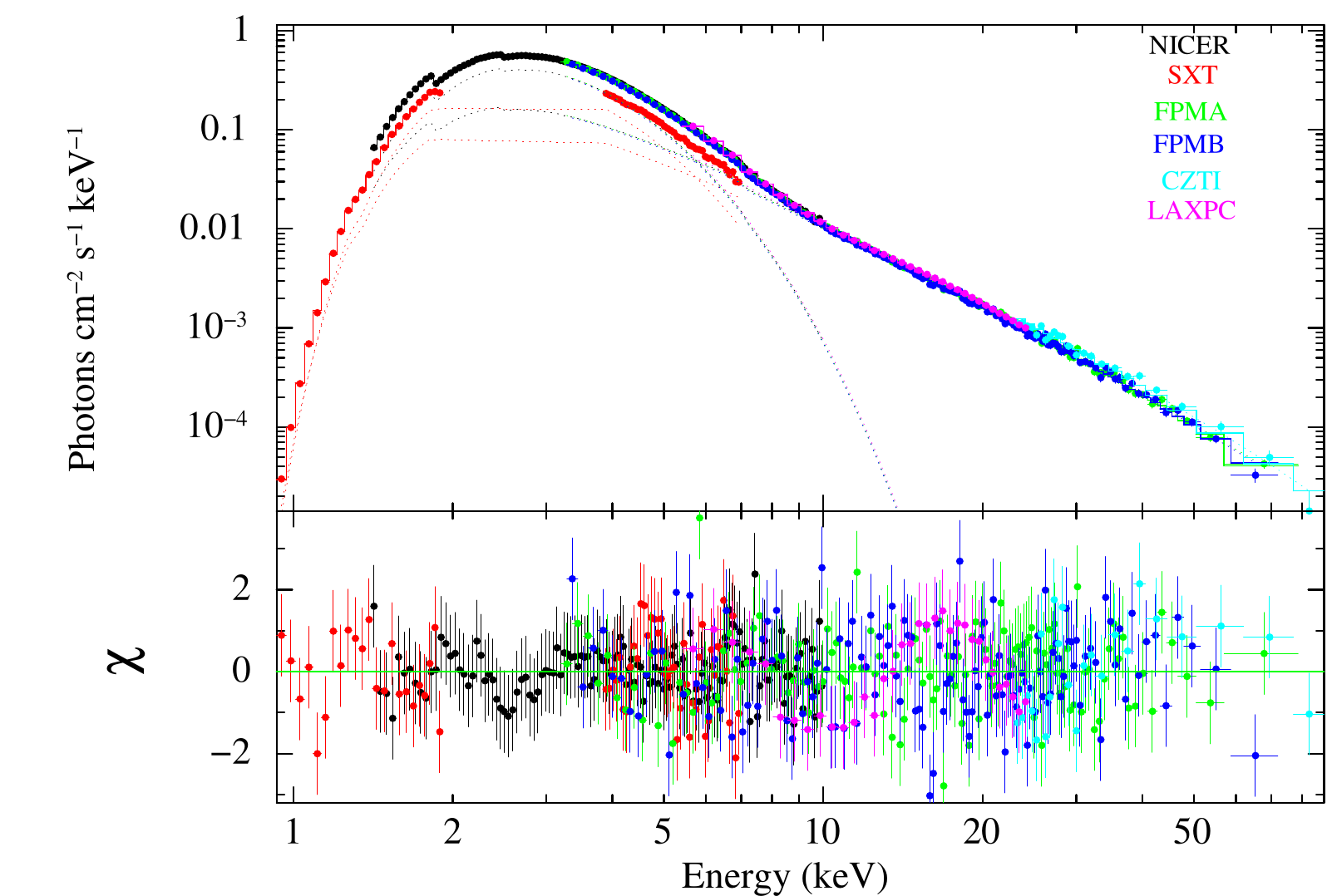}
 \caption{Best-fit broad-band spectra of \src{} with model \texttt{tbabs $\times$ (diskbb + relxillCp)}. The bottom panel represents the residuals obtained from the spectral fit.}
\label{fig:spec_reflection}
\end{figure}
\section{Results}
\label{sec:results}
We performed a detailed timing and spectral analysis of \src{} using X-ray and radio observations carried out during its 2024 outburst. Figure~\ref{fig:BAT} presents the \swift{}/BAT light curve of the source in the 15--50 keV range (top panel) together with the 1.28 GHz radio flux density measured by \meerkat{} (bottom panel). The epochs of the \astrosat, \nicer, \swift, \nustar, and \meerkat{} observations are marked by the magenta, black, green, and blue arrows, respectively. Based on the \insight~HXMT observations, it is presumed that the contribution from the nearby source Cir~X-1 was limited to the soft X-ray band, as it was in the soft state during the observation. Consequently, contamination from the nearby source is expected to be negligible in the hard X-ray band \citep{Peng2024, Chatterjee2025}. 

The radio emission from \src~ is found to be closely associated with the evolution of the hard X-ray outburst as observed with \swift{}/BAT. The radio emission exhibited substantial variability throughout the outburst. It brightened rapidly during the rising phase, reached a maximum during the early decay of the X-ray outburst, and then gradually faded. A subsequent re-brightening was observed during the later stages of the outburst, coincident with a renewed increase in the hard X-ray flux traced by \swift{}/BAT. This close correspondence between the radio and hard X-ray evolution suggests a strong coupling between the jet activity and the accretion flow.
\subsection{X-ray Timing Analysis}
\label{sec:timing}
To investigate the rapid X-ray variability, we extracted the power density spectrum (PDS) from the \textit{AstroSat} \lxp{}20 event data in the 3--25~keV band over the frequency range 0.05--30~Hz using the \texttt{LAXPCSoftware} tasks \texttt{laxpc\_find\_freqlag} and \texttt{laxpc\_rebin\_power}. The resulting PDS from this routine is rms-normalized and corrected for dead-time-modified Poisson noise and background. The rebinned PDS was fitted in \texttt{XSPEC} using multiple Lorentzian components, with the dead-time-corrected Poisson noise level included as the background. The best-fit PDS in the 3-25 keV range reveals a prominent low-frequency QPO at a centroid frequency of $8.07\pm0.04$~Hz with a full width at half maximum (FWHM) of $2.96 ^{+0.14}_{-0.13}$~Hz, corresponding to a quality factor of $Q = 2.73\pm0.14$ ($Q=\nu_{0}/\mathrm{FWHM}$). The integrated fractional rms amplitude of the QPO, derived from the Lorentzian fit, is $5.68^{+0.10}_{-0.09}\%$. The best-fitting model yields a fit statistic of $\chi^{2}/\mathrm{d.o.f.}=242/173$. To further investigate the energy dependence of the QPO, the data were divided into six narrow energy bands (3--5, 5--7, 7--10, 10--12, 12--15, and 15--25~keV). The QPO is detected across both the soft and hard X-ray bands, demonstrating that the feature is intrinsic to \src. 

The best-fit PDS along with the residuals for two different energy ranges, 3-10 keV and 10-25 keV, is shown in Figure~\ref{fig:PDS}, and the corresponding best-fit parameters are summarized in Table~\ref{tab:pds}. The QPO centroid frequencies are measured to be $7.95\pm0.07$ Hz and $8.26\pm0.06$ Hz in the 3--10 keV and 10--25~keV bands, respectively. The corresponding quality factors are $2.47^{+0.20}_{-0.19}$ and $3.68^{+0.66}_{-0.25}$, while the integrated fractional rms amplitudes derived from the Lorentzian fits are $4.84 \pm 0.13\%$ and $13.5^{+0.5}_{-1.1}\%$, respectively. The energy dependence of the intrinsic fractional rms amplitude at the QPO frequency was then computed using the \texttt{laxpc\_find\_freqlag} task over the frequency range 0.05--30~Hz, which estimates the intrinsic fractional rms from the dead-time- and background-corrected Fourier power at the selected QPO frequency ($\sim$8.07~Hz). The intrinsic fractional rms amplitude at the QPO frequency is found to increase from $\sim0.4\%$ to $\sim2.4\%$, as shown in Figure~\ref{fig:frac_rms}. The energy dependence of the intrinsic fractional rms amplitude shows a steady increase from the disk-dominated 3--5 keV band to the power-law-dominated 10--15 keV band, followed by a plateau or marginal decline at higher energies, which may indicate a reduced contribution of the reflection component to variability at higher energies. The timing properties of the detected $\sim8$ Hz QPO, together with the contemporaneous disk-dominated spectrum, are broadly consistent with those of low-frequency QPOs observed during the soft/intermediate state of black hole X-ray binaries \citep{Kalemci2022,Motta2011}.

\subsection{Broadband X-ray Spectroscopy}
\label{sec:spec}
To investigate the broadband spectral properties of \src, we performed a broadband spectral fit to quasi-simultaneous data from \astrosat~ SXT, LAXPC, and CZTI, together with contemporaneous \nicer~ and \nustar~ observations, using {\tt XSPEC} v12.13.0c \citep{Arnaud1996}. To account for cross-calibration uncertainties among the instruments, we included a multiplicative constant component in all spectral models. The constant was fixed at unity for \nicer~ and allowed to vary for the remaining instruments, yielding their relative cross-normalization factors with respect to \nicer. Since launch, the gain of the \sxt{} instrument has drifted by a few tens of eV \citep{Beri2021,Jana2022,Sharma2024}. Therefore, gain corrections were applied to both \sxt{} and LAXPC following the recommendations of the instrument teams \citep{Antia2021,Steiner2024}. To account for a known instrumental feature associated with the Si~K edge\footnote{\url{https://heasarc.gsfc.nasa.gov/docs/nicer/data_analysis/workshops/NICER-CalStatus-Markwardt-2021.pdf}} in the \nicer\ spectra, an absorption edge was included at $\sim$1.8~keV with the optical depth of $\tau=0.2$ \citep{Mandal2023}. Interstellar absorption was modeled using the \texttt{TBabs} model \citep{Wilms2000}. We first explored several simple continuum models, including a single-temperature blackbody (\texttt{bbodyrad}), a multi-color disk blackbody (\texttt{diskbb}; \citealt{Mitsuda1984, Makishima1986}), thermal Comptonization (\texttt{nthComp}; \citealt{Zdziarski1996, Zycki1999}), and a power-law component. All single-component models yielded poor fits with reduced $\chi^{2}_{\nu} \gg 2$.

We then tested more complex continuum descriptions, including \texttt{TBabs}$\times$(\texttt{diskbb} + \texttt{nthComp}) and \texttt{TBabs}$\times$(\texttt{diskbb} + \texttt{powerlaw}). The residuals from all these combinations revealed clear signatures of X-ray reflection, including broad emission features between 5--8~keV, an absorption-like dip near 10~keV, and a Compton hump around 20~keV \citep{Fabian2000, Miller2007}. The reflection features are present regardless of the continuum model. The residuals from fitting the broadband spectra with the \texttt{TBabs}$\times$(\texttt{diskbb} + \texttt{powerlaw}) model are presented in Figure~\ref{fig:residual_plot}. The figure shows excess residuals at $\sim$6.4 keV (Fe~K$_\alpha$ emission line) and $\sim$20 keV (Compton hump). The parameters obtained from the fit are -- a photon index of $\sim$2.6, an inner disk temperature of $\sim$0.9 keV, a disk normalization of $1474_{-22}^{+16}$, and a high reduced~$\chi^2$ of $\sim$1.6. 
\begin{figure*}
\centering
 \includegraphics[width=0.33\linewidth]{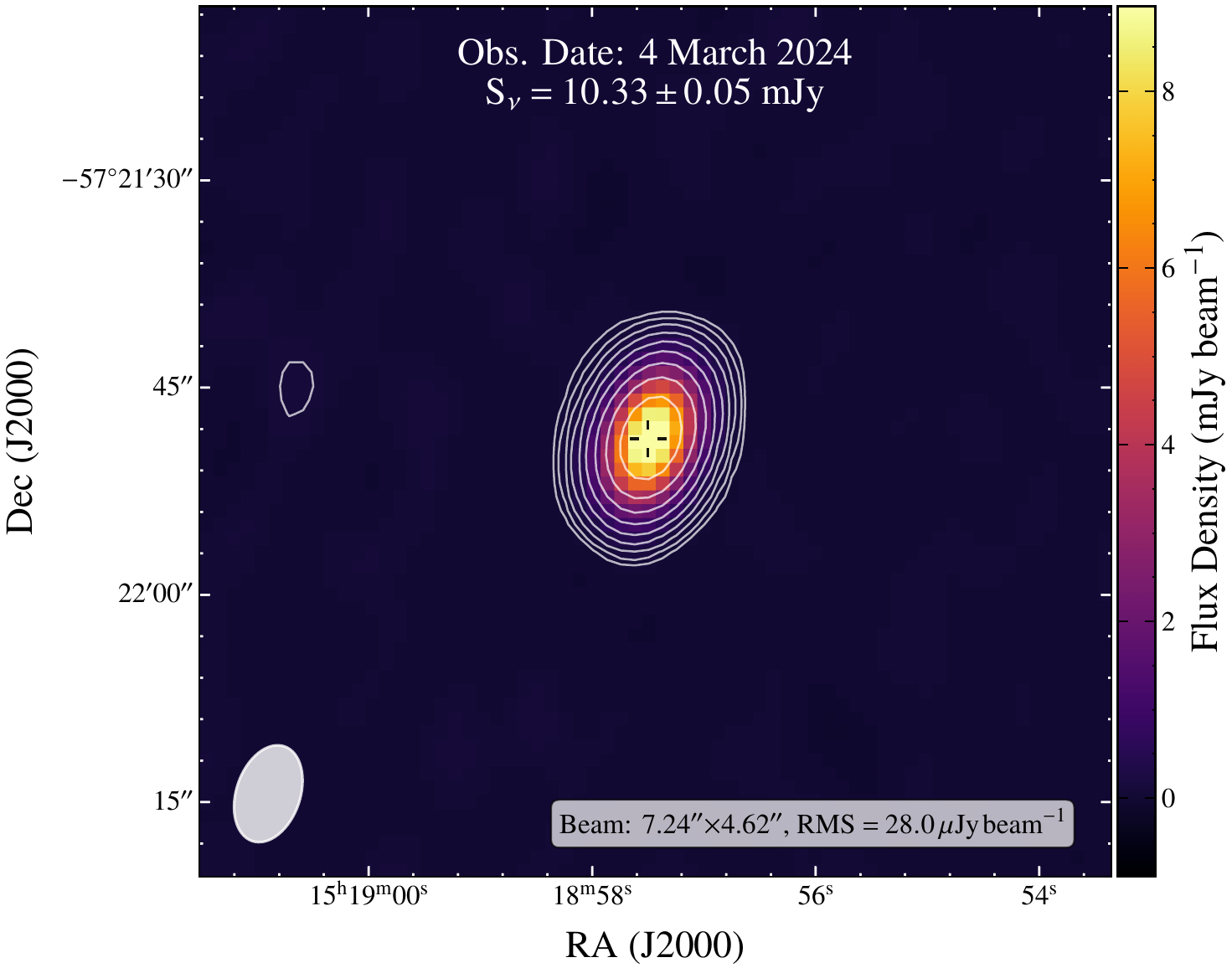}
  \includegraphics[width=0.33\linewidth]{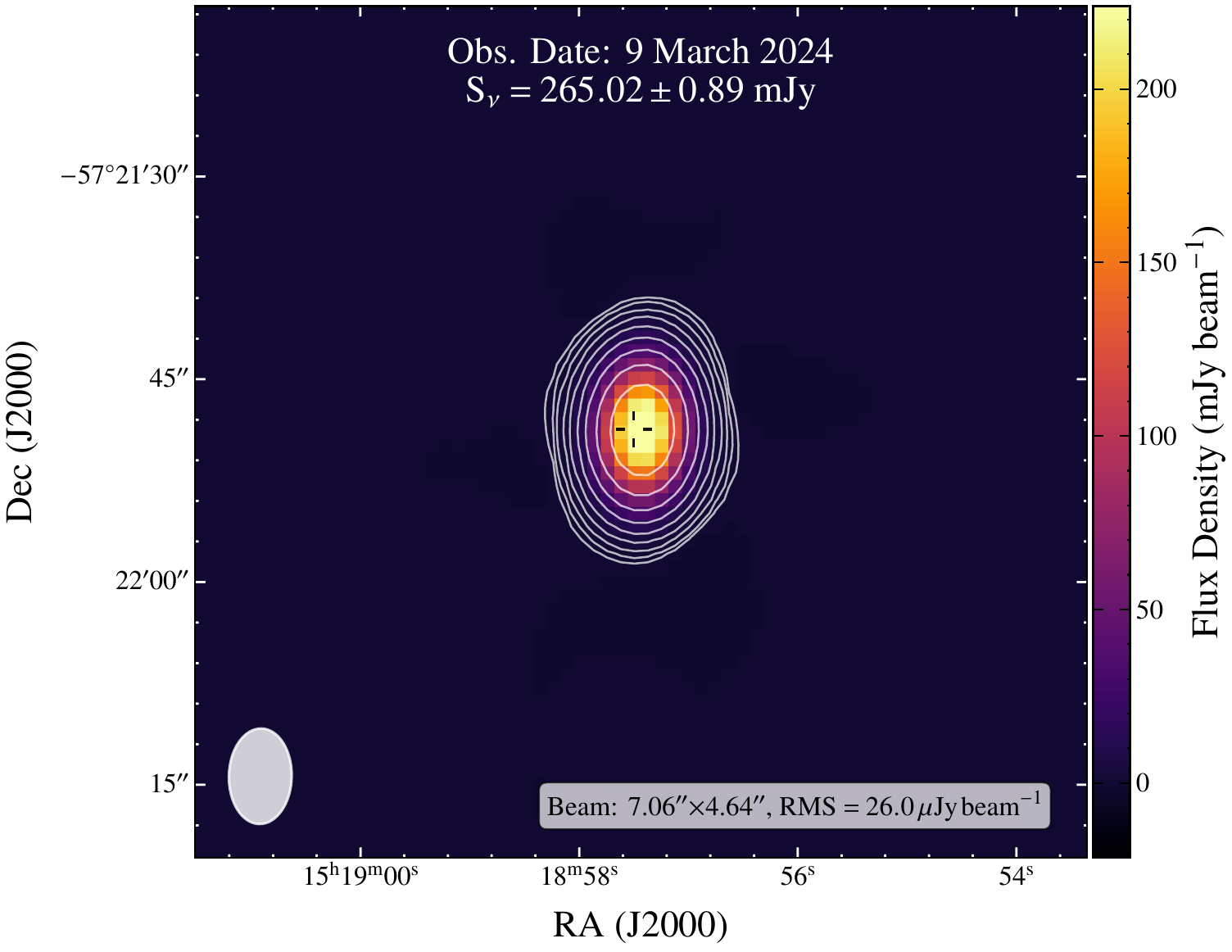}
    \includegraphics[width=0.33\linewidth]{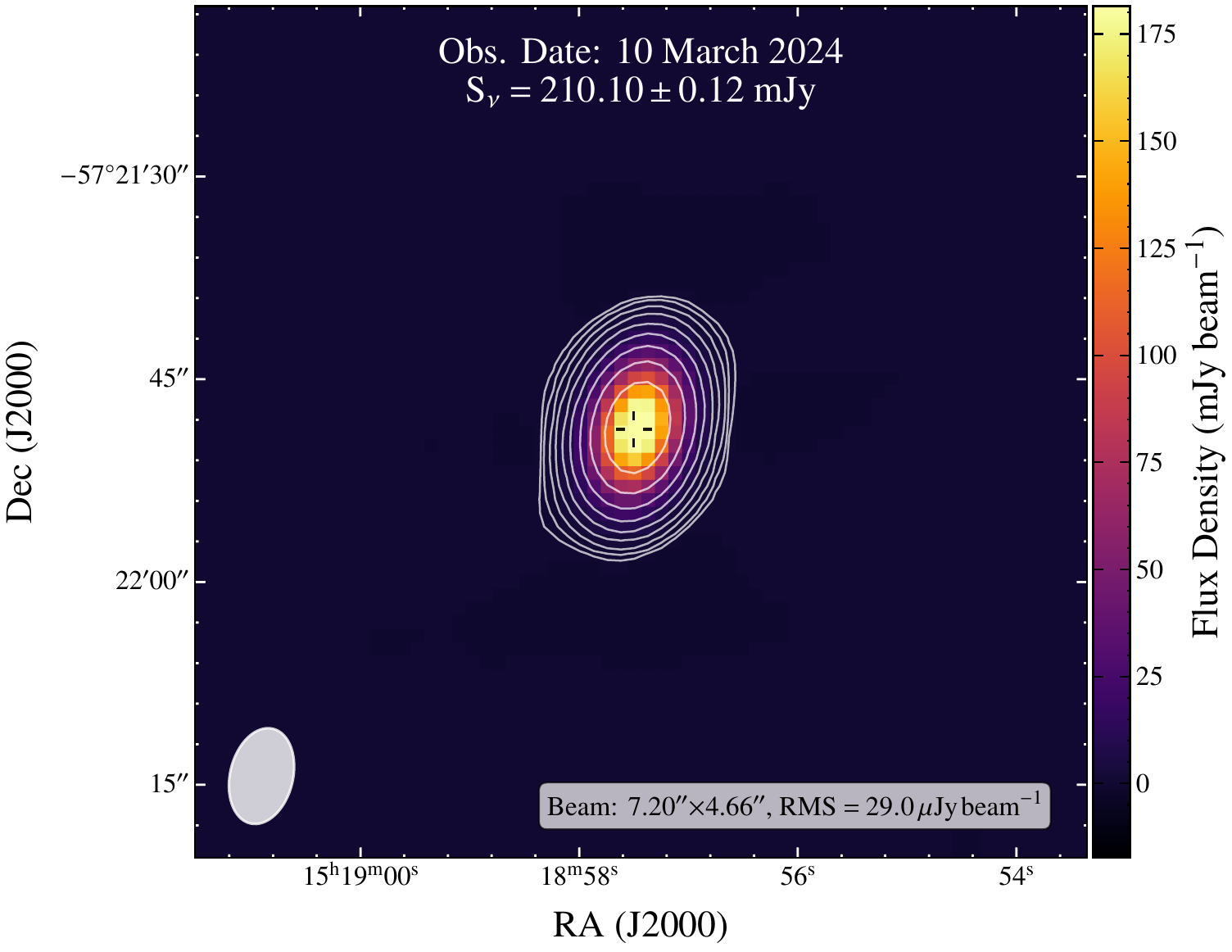}
      \includegraphics[width=0.33\linewidth]{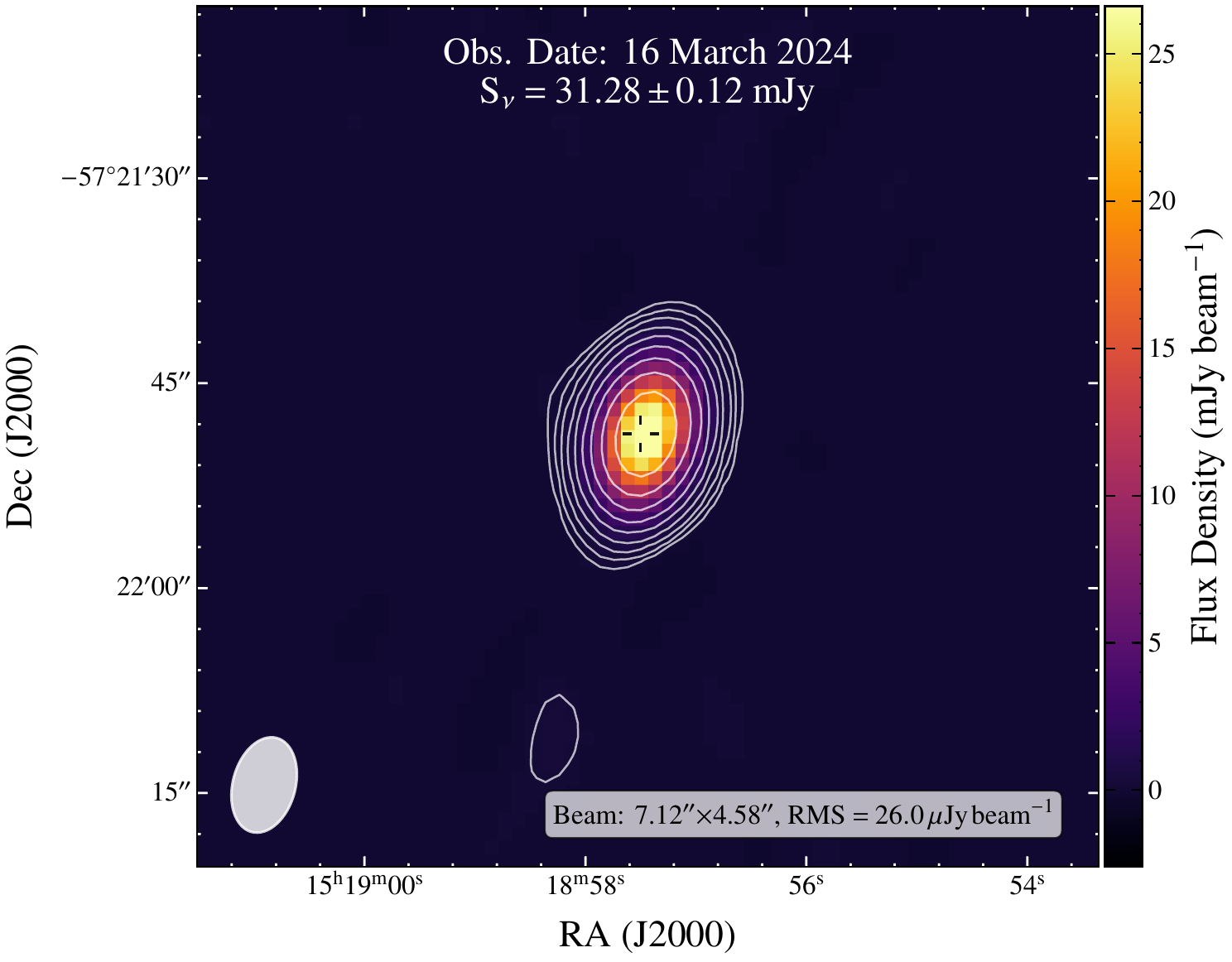}
        \includegraphics[width=0.33\linewidth]{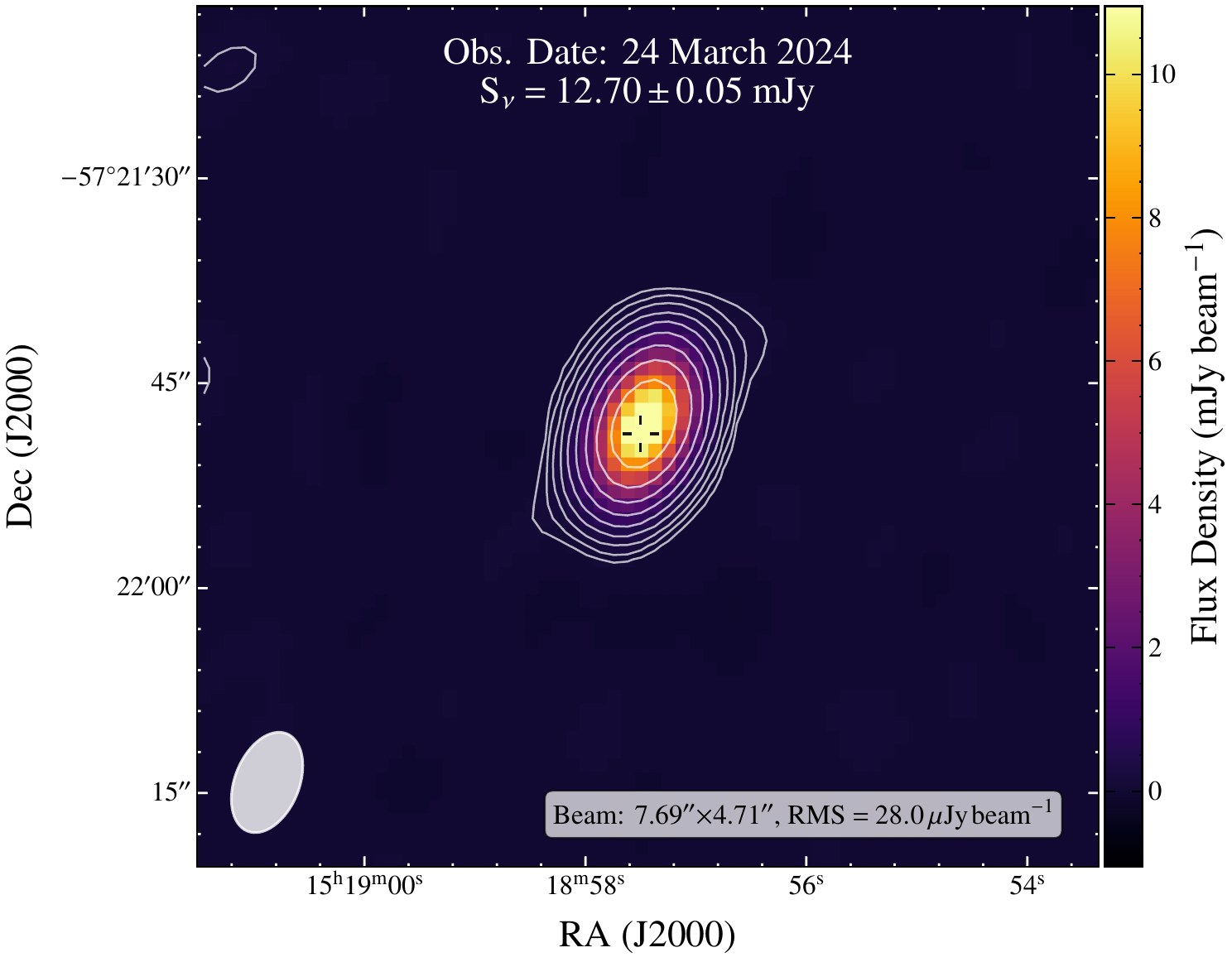}
          \includegraphics[width=0.33\linewidth]{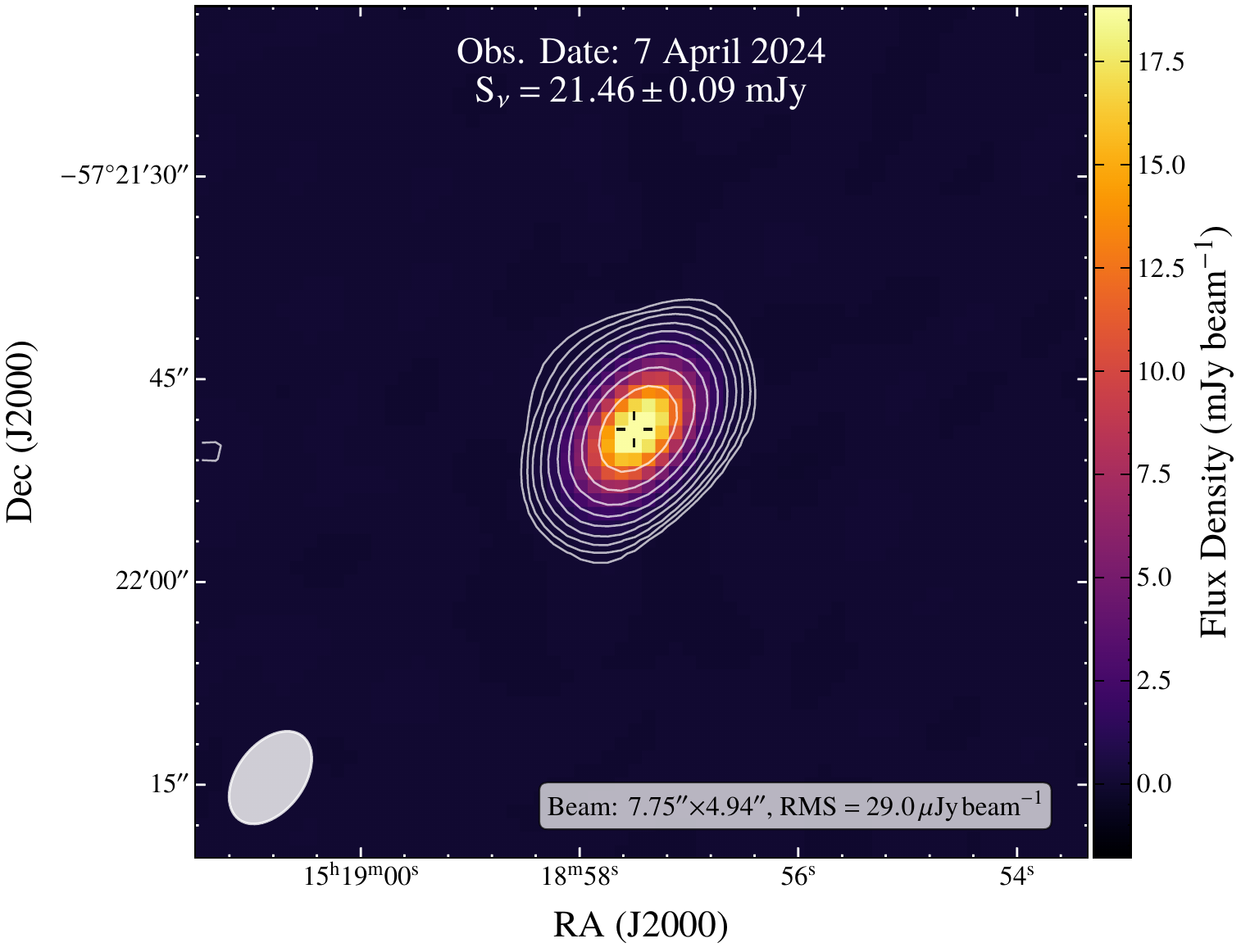}
 \caption{\meerkat\ 1.28 GHz continuum images of the field centered on the black hole X-ray binary \src, obtained at different phases of the outburst, as listed in Table~\ref{tab:obs} and marked in Figure~\ref{fig:BAT}. The synthesized beam is shown in the bottom-left corner of each panel. Contours are plotted at 10 logarithmically spaced levels, starting from $3\sigma$, where $\sigma$ is the local background rms noise. The synthesized beam size and the local background rms of each image are annotated in the lower-right corner of each panel. The black crosshair marks the position of the radio counterpart at $\alpha_{\rm J2000} = 15^{\mathrm h} 18^{\mathrm m} 57.5^{\mathrm s}$ and $\delta_{\rm J2000} = -57^\circ 21' 48.7''$.}
\label{fig:radio_image}
\end{figure*}
To model the reflection features, we employed the relativistic reflection models \texttt{relxill} and \texttt{relxillCp}, both belonging to the \texttt{relxill} family (v2.3; \citealt{Ga14,Da14}). These models self-consistently describe the relativistic reflection spectrum arising from an accretion disk illuminated by a corona. The \texttt{relxill} model assumes a cutoff power-law continuum as the illuminating source, whereas \texttt{relxillCp} uses the thermal Comptonization model \texttt{nthcomp}, providing a more physical description of the coronal emission. In these models, the disk emissivity profile is described by a broken power law characterized by the emissivity indices $q_1$ and $q_2$ and a break radius ($R_{\rm br}$). We adopted a single emissivity profile by setting $q_1=q_2=3$, effectively removing the dependence on $R_{\rm br}$. The reflection models include several physical parameters, such as the photon index ($\Gamma$) of the illuminating continuum, the high-energy cutoff ($E_{\rm cut}$; for \texttt{relxill}), electron temperature ($kT_{\rm e}$; for \texttt{relxillCp}), the inner and outer disk radii ($R_{\rm in}$ and $R_{\rm out}$), disk inclination angle ($i$), dimensionless spin parameter ($a^\ast$), disk ionization parameter ($\log\xi$), and iron abundance ($A_{\rm Fe}$). The outer disk radius was fixed at $R_{\rm out}=400,r_{\rm g}$, where $r_{\rm g}=GM/c^2$, while the inner disk radius was fixed at the innermost stable circular orbit (ISCO). The dimensionless spin parameter was fixed at $a^\ast=0.998$. The hydrogen column density ($N_{\rm H}$), iron abundance, and other continuum and reflection parameters were allowed to vary during the fitting.

We fitted the broadband spectrum with two reflection-based spectral models: M1 = {\tt TBabs $\times$ (diskbb + relxillCp)} and M2 = {\tt TBabs $\times$ (diskbb + relxill)}, together with a multiplicative constant to account for cross-calibration differences between the instruments. The multiplicative constant was fixed at 1 for \nicer~ and left free for the other instruments, yielding best-fit values of 0.70 (SXT), 0.98 (FPMA), 0.95 (FPMB), 1.14 (CZTI), and 1.15 (LAXPC). The line-of-sight absorption due to Galactic neutral hydrogen is considered with the {\tt TBabs} model \citep{Wilms2000} in {\tt XSPEC}. The most recent abundance model, {\tt wilm}, is adopted and incorporated into {\tt XSPEC} for the input abundance in the {\tt TBabs} model. Both models provide statistically acceptable fits to the broadband spectrum, yielding a reduced $\chi^2$ of $\sim$0.96. Figure~\ref{fig:spec_reflection} shows the best-fit broadband spectra of \src{} obtained with model M1. A gain correction was applied to both \sxt{} and \lxp{} to account for residual calibration uncertainties. The gain slope is fixed to 1, and the offset is allowed to vary. The best-fit gain parameters correspond to offsets of $-0.012$ keV for \sxt{} and $-0.7$ keV for \lxp{}, respectively. The residuals displayed in the lower panel indicate that the model adequately reproduces the observed reflection features. 

For model M1, the disk inclination is constrained to be $27.4^{+3.5}_{-4.6},^{\circ}$. The illuminating continuum is characterized by a photon index of $\Gamma = 2.64^{+0.04}_{-0.03}$, while model M2 yields a high-energy cutoff of $E_{\rm cut}=80.2^{+12.4}_{-7.4}$ keV. In the case of model M1, the electron temperature is found to be $kT_{\rm e}=77.4^{+29.9}_{-23.6}$ keV, with a disk density of $\log N = 15.1^{+1.06}_{-0.01}~{\rm cm^{-3}}$. The accretion disk is highly ionized in both models, with an ionization parameter of $\log\xi = 3.68^{+0.18}_{-0.35}~{\rm erg~cm~s^{-1}}$. The iron abundance ($\rm A_{Fe}$) is found to be in the range of $3.0 ^{+0.8}_{-1.1}$, and the reflection fraction is $0.32^{+0.12}_{-0.07}$ from the reflection modeling with spectral model M1. The multi-temperature disk blackbody component ({\tt diskbb}) yields an inner disk temperature of $\sim0.9$ keV with a normalization of $1752.7^{+80.6}_{-66.3}$. The unabsorbed flux in the 0.1--100 keV band is $(8.66\pm0.03)\times10^{-8}~{\rm erg~cm^{-2}~s^{-1}}$, corresponding to a luminosity of $(3.48\pm3.00)\times10^{38}~{\rm erg~s^{-1}}$ for an assumed source distance of $5.8\pm2.5$ kpc \citep{Peng2024}. The best-fit parameters for both models are summarized in Table~\ref{tab:fitstat2}.  The accretion disc parameters obtained from two different combinations of reflection models (M1 and M2) are consistent (within errors). Parameter uncertainties are estimated using the Markov Chain Monte Carlo ({\tt MCMC}) method. We employed the Goodman--Weare algorithm \citep{Good2010} with the reflection models ({\tt relxill/relxillCp}), generating chains of length 500000 using 10 walkers. The first 50000 steps were discarded as burn-in before deriving the parameter confidence intervals. 

In addition, we analyzed three \textit{Swift}/XRT observations listed in Table~\ref{tab:obs} to extend the X-ray coverage and obtain the 1–10~keV flux required for the radio–X-ray correlation analysis. The XRT spectra were fitted using the same continuum model, {\tt TBabs $\times$ (diskbb + po)}, which provided satisfactory fits with reduced $\chi^2$ of 1.04--1.16. The best-fit photon index varied between --0.5 and 2.7, while the inner-disc temperature ranged from 0.65 to 1.17~keV. The hydrogen column density is found to be in the range of $(6.7-7.2)\times10^{22}$ cm$^{-2}$. The corresponding 1–10~keV unabsorbed flux varied in the range of $1.0-3.8\times10^{-8}$ ~erg~cm$^{-2}$~s$^{-1}$. These flux measurements were subsequently used to investigate the radio–X-ray correlation of the source.
\begin{figure}
    \includegraphics[width=0.475\textwidth]{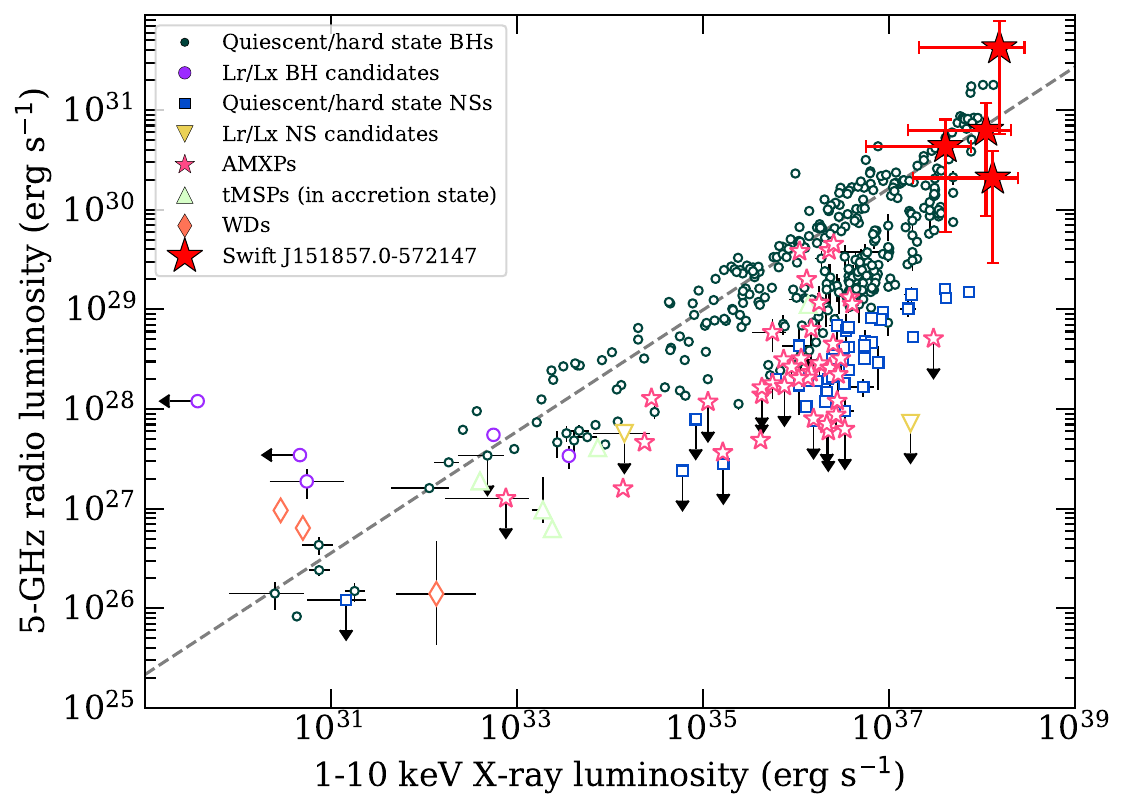}
    \caption{Radio luminosity ($L_{\rm R}$) at 5~GHz versus X-ray luminosity ($L_{\rm X}$) in the 1-10 keV range for accreting compact objects, adopted from \citet{Bahramian2022}. The dotted black line indicates the best-fit relation for black holes \citep{Gallo2006}.  The red asterisk denotes the position of \src\ in the $L_{\rm R}$--$L_{\rm X}$ plane, based on quasi-simultaneous \meerkat\ and X-ray observations from its 2024 outburst. The source is consistent with the black hole X-ray binary population in the $L_{\rm R}$–$L_{\rm X}$ plane.}
    \label{fig:lx_lr}
\end{figure}
\begin{table}
\scriptsize
	\centering
	\caption{Best-fit parameters obtained from broadband spectral fitting in the energy range 0.9--90 keV using the reflection model combinations. 
    The best fits are obtained with the reflection models M1: \texttt{tbabs $\times$ (diskbb + relxillCp)} and M2: \texttt{tbabs $\times$ (diskbb + relxill)}.}
	\label{tab:fitstat2}
    \setlength{\tabcolsep}{2.7pt}   
	\begin{tabular}{llcc} 
\hline
 Model & Parameters	&   M1& M2	\\	
\hline																			
\texttt{tbabs} &N$_\text{H}$ ($10^{22}$ cm$^{-2}$) 	& $6.32_{-0.04}^{+0.05}$  & $6.29_{-0.03}^{+0.05}$ \\		
\hline
\texttt{diskbb} & $T_{in}$ (keV) &  $0.90_{-0.007}^{+0.006}$ & $0.89_{-0.005}^{+0.006}$    \\

& Norm$_\mathrm{bb}$ & $1752.0_{-66.3}^{+80.6}$ &$1778.0_{-64.8}^{+56.6}$ \\ 
\hline
\texttt{relxill/} & Photon index ($\Gamma$)	 & $2.64_{-0.03}^{+0.04}$ & $2.58_{-0.02}^{+0.04}$ \\ 

\texttt{relxillCp}& kT$_{\rm e}$ (keV)  &$77.4_{-23.6}^{+29.9}$ &  $-$\\ 

& $E_{\mathrm{cut}}^e$ (keV)	&  	$-$ &  $80.2_{-7.4}^{+12.4}$ \\ 

& Incl (deg) 	& $27.4_{-4.6}^{+3.5}$ 	&	$27.2_{-1.7}^{+2.6}$ \\ 

& $R_{\mathrm{in}}$ ($R_{\mathrm{ISCO}})$	& $-1^c$ 	&	 $-1^c$  \\
& $R_{\mathrm{out}}$ ($R_{\mathrm{g}})$	& $400^c$ 	&	 $400^c$  \\

& $A_{\mathrm{Fe}}$ &	 $3.0_{-1.1}^{+0.8}$ 	&	$4.8_{-2.0}^{+0.6}$  \\

&q1=q2	 & $3^c$ 	&	 $3^c$  \\ 

&$a^\ast$	 &	 $0.998^c$ 	&	 $0.998^c$  \\ 

& $\log \xi\;(\mathrm{erg~cm~s}^{-1})$ & $3.68_{-0.35}^{+0.18}$ & $3.82_{-0.14}^{+0.16}$ \\

& $\log\text{N}$ (cm$^{-3}$) 		& $15.1_{-0.01}^{+1.06}$  	&	 \\

& $\mathrm{Refl}_{\mathrm{frac}}$	&	$0.32_{-0.07}^{+0.12}$ 	&	$0.25_{-0.03}^{+0.07}$  \\

& $\mathrm {Norm}_{\mathrm{Refl}}$ ($\times10^{-2})$  & $5.40_{-0.4}^{+0.5}$ 	&	$5.80_{-0.3}^{+0.5}$ \\
\hline 
& Total Flux$^a$  & $8.66 \pm 0.03 $ 	&	$8.30 \pm 0.02$  \\
 & Luminosity$^b$ 	  	& $3.48 \pm 3.00$ 	& $3.34 \pm 2.88$  \\
\hline                   															
 & Reduced $\chi^2$ (dof)   & 0.96 (627)        &  0.95(628)   \\
\hline		
\multicolumn{3}{l}{$^a$ : Unabsorbed flux (0.1-100 keV) in the units of $10^{-8}$ \erg.}\\
\multicolumn{3}{l}{$^b$ : in the units of $10^{38}$ \lum, for a distance of $5.8 \pm 2.5$ kpc.}\\
\multicolumn{3}{l}{$^c$ : Frozen parameters.}\\
\multicolumn{3}{l}{$^e$ : { $E_{\rm cut}$, high-energy cutoff of illuminating power-law continuum.}}\\ 
\end{tabular}
\end{table} 

\subsection{Radio/X-ray correlation}\label{sec:radio}
Following the discovery of the X-ray transient \src, radio monitoring observations were carried out with \meerkat\ to investigate its properties across multiple wavebands. The source was observed regularly during the 2024 outburst. We used radio observations obtained between 2024 March 4 and April 7 at a central frequency of 1.28 GHz in continuum mode. The radio flux density evolved significantly during the outburst, as shown in Figure~\ref{fig:BAT}. It was initially $10.3 \pm 0.05$ mJy on 2024 March 4, during the rising phase of the hard X-ray outburst, and increased rapidly over the following few days to a peak of $265 \pm 0.9$ mJy on 2024 March 9. Thereafter, the flux density declined steadily to $12.7 \pm 0.05$ mJy by 2024 March 24. A subsequent re-brightening was observed, with the radio flux density increasing to $21.46 \pm 0.09$ mJy on 2024 April 7. This evolution broadly followed the changes in the \swift{}/BAT hard X-ray light curve, with a secondary radio brightening coinciding with a renewed increase in the hard X-ray flux. Figure~\ref{fig:radio_image} presents the 1.28 GHz \meerkat{} continuum images of \src{} obtained during the monitoring campaign. One of the \meerkat{} continuum images of \src, obtained during a quasi-simultaneous \astrosat{} observation, showing a radio point source at the nominal position of \src{} (RA = $15^{\mathrm h} 18^{\mathrm m} 57.5^{\mathrm s}$, Dec = $57^\circ\,21'\,48.7''$) with a measured flux density of $31.28\pm0.12$ mJy on 2024 March 16 (MJD 60385.06). Though the radio continuum flux varied significantly during the X-ray outburst, a compact radio counterpart is clearly detected at the source position in all observing epochs.  

 The quasi-simultaneous \astrosat\ and \meerkat\ observations performed on 2024 March 16 provide the first opportunity to investigate the X-ray/radio correlation of the newly discovered black hole X-ray binary \src. {In addition, we used three quasi-simultaneous X-ray observations performed with \swift/XRT on MJD 60373, 60379, and 60407.} Such studies are important for understanding the coupling between accretion disk and jet emission and for determining the location of the source in the radio-X-ray luminosity plane. For this analysis, we estimated the X-ray luminosity in the 1--10 keV band and the radio luminosity at 5 GHz. The latter was derived from the observed \meerkat\ 1.28 GHz flux density, assuming a flat radio spectrum ($S_{\nu}\propto\nu^{0}$). Adopting a source distance of $5.8\pm2.5$ kpc \citep{Peng2024}, we obtained a 5 GHz radio luminosity of $L_{\rm R}=(6.29\pm5.42)\times10^{30}$ erg\,s$^{-1}$. The unabsorbed 1--10 keV X-ray flux frorm \astrosat~ observation was estimated to be $(2.72\pm0.02)\times10^{-8}$ erg cm$^{-2}$ s$^{-1}$, corresponding to an X-ray luminosity of $(1.10\pm0.94)\times10^{38}$ erg s$^{-1}$. The X-ray flux in the 1–10 keV band from the XRT observations on MJD 60373, 60379, and 60407 are $(3.22\pm0.02)\times10^{-8}$, $(3.80\pm0.02)\times10^{-8}$, and $(1.00\pm0.01)\times10^{-8}$ erg cm$^{-2}$ s$^{-1}$, respectively. The corresponding X-ray luminosities, assuming a source distance of $5.8\pm2.5$ kpc, are estimated to be $(1.30\pm1.12)\times10^{38}$, $(1.53\pm1.32)\times10^{38}$, and $(4.03\pm3.47)\times10^{37}$ erg s$^{-1}$, respectively. On the corresponding days, the radio flux densities are measured to be $10.33\pm0.05$ mJy, $210.10\pm0.12$ mJy, and $21.46\pm0.09$ mJy, respectively. The 5~GHz radio luminosity is also estimated from the radio flux densities, assuming a flat radio spectrum and a source distance of $5.8\pm2.5$ kpc, yielding $L_{\rm R}=(2.08\pm1.79)\times10^{30}$, $(4.23\pm3.65)\times10^{31}$, and $(4.32\pm3.72)\times10^{30}$ erg~s$^{-1}$, respectively. Figure~\ref{fig:lx_lr} shows the position of \src~ in the $L_{\rm X}$--$L_{\rm R}$ plane together with the well-established tracks followed by black hole and neutron star X-ray binaries. The uncertainty in the source distance contributes to the large uncertainties in the radio and X-ray luminosities. Within these uncertainties, \src\ is consistent with the distribution of known black hole X-ray binaries, further supporting its classification as a black hole.
\section{Discussion \& Conclusions}
\label{sec:discon}
We perform a broadband timing and spectral study of the newly discovered black hole X-ray binary \src~ using multiwavelength data from \astrosat, \nicer, \swift, \nustar, and \meerkat. The \astrosat~ observation was obtained close to the first epoch (2024 March 18 (MJD 60387)) of the 15 \nicer~ observations. Based on the hardness-ratio evolution throughout the 15 \nicer~ observations during 2024 March 18 -- 2024 May 03, \citet{Peng2024} reported that the source remained in the soft state during the above duration, where the accretion disk dominates the emission and extends close to the ISCO. The temporal evolution of the radio and X-ray emission suggests a possible delay of the radio emission with respect to the X-rays (Fig.~\ref{fig:BAT}). Such delayed radio variability has been reported in several black-hole X-ray binaries and investigated through cross-correlation analyses of radio and X-ray light curves (e.g., \citealt{Tetarenko2019,Shapopi2021,You2024}). This delay may arise from the response of the jet to fluctuations in the inner accretion flow, including the propagation of these fluctuations to the radio-emitting region and the time required for the synchrotron-emitting plasma to become optically thin. In the present case, however, the sparse radio sampling prevents a quantitative determination of the lag. The initial two {\it MeerKAT} observations are obtained on MJD~60373.08 and 60378.11, with the radio flux increasing from $10.33\pm0.05$ to $265.02\pm0.89$~mJy, respectively, with no intermediate radio observation available. Thus, the exact epoch of the radio peak, and hence the radio--X-ray lag, cannot be constrained. We therefore interpret the apparent delay only as a possible indication of a delayed jet response rather than as a measured lag. Denser radio monitoring during future outbursts would enable a cross-correlation analysis and provide stronger constraints on the disk--jet coupling in this source.

In this context, the X-ray timing properties provide further insight into the variability of the accretion flow during the observed accretion state. Timing analysis of the \textit{AstroSat} LAXPC20 data reveals the presence of a prominent $\sim8$ Hz QPO in the 3--25 keV band, which is also detected in both the 3--10 and 10--25 keV energy bands. The QPO detected in the 3--25 keV range PDS has a centroid frequency of $8.07 \pm 0.04$ Hz, a quality factor of $Q = 2.73 \pm 0.14$, and an integrated fractional rms amplitude of $5.68^{+0.10}_{-0.09}\%$. We further investigated the energy dependence of the QPO. We found that both the centroid frequency and the integrated fractional rms amplitude increase from the 3--10 to the 10--25~keV band, indicating that the QPO variability becomes stronger at higher energies. The relatively broad profile, low quality factor ($Q\sim2.5$), moderate QPO fractional rms ($\sim$4.8\%), and relatively low total fractional rms ($\sim$8.1\%) in the 3--10 keV band are broadly consistent with the properties of type-A low-frequency QPOs observed in black hole X-ray binaries, although both the QPO fractional rms and total fractional rms are somewhat higher than the typical values reported for canonical type-A QPOs ($\lesssim$3\% and $\lesssim$5\%, respectively) \citep{Casella2004,Casella2005, Motta2011,Zhang2023}.

Type-A QPOs are generally observed around 6-8 Hz and characterized by broad profiles with low quality factors ($Q\lesssim3$), low fractional rms amplitudes (typically $\lesssim3\%$), and weak red-noise components, whereas neither a subharmonic nor a second harmonic was present \citep{Casella2005,Motta2011}. Type-A QPOs are less frequently observed and rarely studied, and no comprehensive model has been proposed for type-A QPOs. Such QPOs have been reported in several black hole X-ray binaries, including  GS~1124--684 \citep{Miyamoto1993}, GX~339--4 \citep{Motta2011, Nespoli2003}, MAXI~J1348--630 \citep{Zhang2023}, and are commonly observed during soft/intermediate states \citep{Casella2005,Motta2011,Zhang2023}. In contrast, Type-C QPOs are generally narrower and stronger, with higher $Q$ and fractional rms, and are typically accompanied by strong flat-top noise \citep{Casella2005,Motta2011}. For comparison, \citet{Chatterjee2025} reported Type-C QPOs during the 2024 outburst of the same source, with the QPO frequency evolving from $\sim$3.2 Hz to $\sim$9 Hz across their observing period. The QPOs generally showed relatively high quality factors and were detected over a broad energy range, extending up to $\sim$48 keV. Their spectral analysis indicated that the source was initially in an intermediate state and subsequently evolved toward the soft state \citep{Chatterjee2025}. This interpretation is also supported by the spectral-state evolution reported in previous studies. In particular, \citet{Peng2024} constructed a \nicer~ hardness–intensity diagram using the $4$--$10$ keV/$1$--$4$ keV hardness ratio and found that the source remained in the soft state during almost the entire \nicer~ monitoring period (MJD 60387.5--60433.0). Notably, our {\it AstroSat} observation preceded the \nicer~ soft-state observations by only $\sim$1.7 days, supporting a soft/intermediate-state interpretation of our observation \citep{Peng2024}. The \nicer~ HID presented by \citet{Song2026} independently places the epoch of our {\it AstroSat} observation (MJD~60386) on the soft/intermediate state branch of the outburst track (see Figure~1b of \citet{Song2026}), further supporting this interpretation.

To investigate the broadband spectral properties of the source, we performed a joint spectral analysis of the quasi-simultaneous \astrosat, \nicer, and \nustar~ observations covering the 0.9--90 keV energy range. The broadband spectrum exhibits clear reflection signatures, including a broadened Fe~K\(\alpha\) emission line and a Compton hump. These features were modeled using the relativistic reflection models {\tt relxill} and {\tt relxillCp}. The best-fit spectral parameters indicate a steep photon index of \(\Gamma \approx 2.6\) and an electron temperature of \(kT_{\rm e} \approx 54-107\) keV. The disk inclination, density, and ionization parameter are constrained to be \(\sim 23-31^\circ\), \(\log n_{\rm e} \approx 15-16\) cm$^{-3}$, and \(\log \xi \approx 3.3-4.0\) erg cm s$^{-1}$, respectively. These parameters are in good agreement with those reported by \citet{Peng2024} from the near-simultaneous \insight~HXMT observation, suggesting a consistent accretion geometry and disk properties across different instruments. The broadband spectral modeling reveals relativistic reflection from a moderately dense, highly ionized accretion disk with a supersolar iron abundance, consistent with other black hole X-ray binaries such as Cyg~X-1 \citep{Parker2015}, MAXI~J1820+070 \citep{You2021}, GX~339-4 \citep{Furst2015}. The reflection spectral modeling further suggests a moderately inclined system, with the accretion disk extending close to the compact object during the observation. The reflection fraction is relatively small ($\sim$0.25-0.45), suggesting that the accretion disk intercepts only a limited fraction of the coronal emission. This is further supported by the low normalization of the reflection component, indicating that the reflected emission contributes only a modest fraction of the total X-ray flux. To further characterize the spectral state, we estimated the unabsorbed 2--20~keV flux, where the \texttt{diskbb} component contributes $\sim63\%$ ($1.0\times10^{-8}$~erg~cm$^{-2}$~s$^{-1}$) of the total flux ($1.58\times10^{-8}$~erg~cm$^{-2}$~s$^{-1}$), while the \texttt{relxillCp} component contributes the remaining $\sim37\%$ ($6.51\times10^{-9}$~erg~cm$^{-2}$~s$^{-1}$). However, the disk fraction is below the $75\%$ threshold commonly adopted for the canonical thermal state \citep{rem06}, the steep photon index ($\Gamma = 2.59$--$2.65$), and the dominant thermal disk ($kT_{\rm in}\sim0.9$~keV) indicate that the source was in a disk-dominated, soft/intermediate spectral state.

The inner disk radius is estimated using the normalization of the {\tt diskbb} model and the best-fit disk inclination angle from the reflection model. From the \texttt{diskbb} normalization, we have

\begin{equation}
N_{\rm diskbb}=\left(\frac{R_{\rm in}}{D_{10}}\right)^2 \cos\theta,
\end{equation}

where $R_{\rm in}$ is the apparent inner disk radius, $D_{10}$ is the source distance in units of 10 kpc, and $\theta$ is the disk inclination angle. Assuming an inclination angle of $\sim28^\circ$ and a best-fit disk normalization of $N_{\rm diskbb}\sim1750$, the inner disk radius can be expressed as

\begin{equation*}
R_{\rm in} = 44.5\,D_{10}~{\rm km}.
\end{equation*}

The uncertainty in the source distance prevents a tighter constraint on the disk radius, as $R_{\rm in}$ scales linearly with $D_{10}$. Adopting a representative distance of $\sim$8 kpc ($D_{10}=0.8$), we obtain an apparent inner disk radius of $R_{\rm in} \simeq 35.6~{\rm km}$.

Applying the standard correction factor ($\kappa^2\xi \simeq 1.19$) to account for spectral hardening and the inner boundary condition, the corresponding physical inner disk radius becomes

\begin{equation*}
R_{\rm in,true} = 53.0\,D_{10}~{\rm km}
\end{equation*}
or
\begin{equation*}
R_{\rm in,true} \simeq 42.4~{\rm km}
\end{equation*}

for a distance of $\sim$8 kpc. For a $10\,M_\odot$ black hole, the inferred $R_{\rm in,true}\simeq42.4$ km corresponds to $\sim2.9 R_g$, where $R_g=GM/c^2\simeq14.8$ km. This indicates that the optically thick disk extends close to the black hole. The inner disk radius ($\sim2.9 R_g$) is considerably smaller than the inner disk radius during the canonical hard state, typically $\sim$10-40$R_g$ \citep{Gierlinski2008, Basak2017, Wang2018}, although the inferred radius depends on the source, luminosity, and modelling assumptions. The estimated inner disk radius during our observation is consistent with a disk extending close to the ISCO, which supports the soft spectral state. The result is also broadly consistent with the inner disk radius reported from the \insight\textit{-HXMT} observations \citep{Peng2024}. 

The evolution of the inner disk radius is closely related to the spectral state of black holes. A recent \nicer~ study of Swift~J1727.8--1613 reports that the inner disk radius was close to the ISCO in the soft spectral state and increased to $\lesssim8 R_{\rm ISCO}$ when the source evolved to a hard spectral state \citep{Konig2026}. In GX~339--4, reflection spectroscopy found the inner disk radius to remain below $\sim9 R_g$ throughout the bright intermediate-state transition \citep{Sridhar2020}. In the case of Cyg~X-1, the inner disk radius is estimated to be $R_{\rm in}\sim13$--$20 R_g$ in the hard state \citep{Basak2017}. We note, however, that the absolute value of $R_{\rm in}$ is subject to systematic uncertainties associated with the black-hole mass, spin, distance, inclination, spectral hardening (color-correction) factor, and the adopted spectral model \citep{Konig2026}. These results, together with the systematic study of black hole X-ray binaries by \citet{Dunn2011}, illustrate the general trend of a relatively stable and smaller $R_{\rm in}$ in the disk-dominated/soft state, while deviations from a constant inner radius become more prominent toward the intermediate and hard states.

The radio emission exhibited pronounced variability throughout the 2024 outburst and broadly tracked the evolution of the hard X-ray emission observed by \swift{}/BAT. In particular, the initial increase in the radio flux density, followed by its gradual decline and subsequent re-brightening accompanying the renewed hard X-ray activity (MJD 60407), is consistent with a close coupling between the accretion flow and the radio-emitting outflow. A dramatic increase in the radio flux density was observed with \meerkat{} at 1.28~GHz, rising from 10.3~mJy on 2024 March 4 to 265~mJy on 2024 March 9, indicating the onset of strong radio activity. Consistent with this behavior, the Australia Telescope Compact Array ({\it ATCA}) observed the source on 2024 March 9 between 10:35:10 and 11:06:20~UT (MJD~60378.45) and reported a bright radio flare with flux densities of $1.70 \pm 0.17$~Jy at 5.5~GHz and $1.55 \pm 0.16$~Jy at 9~GHz \citep{car24}. Such intense radio flaring is commonly associated with transient relativistic ejecta produced during spectral state transitions in Galactic black hole X-ray binaries citep{Fender2004,Fender2009, Hughes2025}.
 
Extrapolating the spectral index measured from the quasi-simultaneous {\it ATCA} observations at 5.5 and 9~GHz ($\alpha \approx -0.19$, where $S_\nu \propto \nu^\alpha$), obtained approximately 6~h after the end of the \meerkat\ observation, to the \meerkat\ observing frequency predicts a flux density of $\sim2.2$~Jy at 1.28~GHz. This is nearly an order of magnitude higher than the estimated \meerkat\ flux density of $265.02 \pm 0.89$~mJy. Assuming negligible spectral evolution over the intervening $\sim6$~h, this discrepancy indicates that a single power-law spectrum cannot describe the radio emission over the 1.28--9~GHz range. Instead, the observations may indicate that the source remained partially synchrotron self-absorbed at L-band, with the spectral turnover likely located between 1.28 and 5.5~GHz. However, intrinsic spectral evolution during the $\sim6$~h separating the \meerkat\ and {\it ATCA} observations may also have contributed to the observed discrepancy. One possible interpretation is that the radio emission originated from a recently launched jet ejecta whose expanding synchrotron photosphere had not yet become optically thin at 1.28~GHz.

Independent lower-frequency monitoring with {\it ASKAP/VAST} at 887.5~MHz \citep{Anumarlapudi24} provides additional insight into the radio evolution. The source was not detected prior to the outburst, with a $5\sigma$ upper limit of $<2.5$~mJy on 2024 February 21 (MJD~60361.82; local image rms of 0.52~mJy), consistent with a radio-faint quiescent state. Following the bright \meerkat/{\it ATCA} flare, VAST measured flux densities of $46.67 \pm 0.35$~mJy on 2024 March 11 (MJD~60380.78) and $7.14 \pm 0.33$~mJy on March 27 (MJD~60396.73), tracing the continued decline of the radio emission (bottom panel of Figure~\ref{fig:BAT}). The evolution of radio flux densities, estimated from \meerkat\ and {\it ASKAP/VAST} observations, closely follows the \swift/BAT hard X-ray flux during the outburst. As the radio observations of \src~ with \meerkat, {\it ATCA}, and {\it ASKAP/VAST} are not simultaneous, and the source was evolving rapidly, a direct comparison of the radio spectra across these epochs is not reliable.

More notably, {\it ASKAP/VAST} independently captures the subsequent radio re-brightening also observed with \meerkat. The flux density increased to $17.55 \pm 0.33$~mJy on 2024 April 12 (MJD~60412.75), following the \meerkat\ measurement of $21.46 \pm 0.09$~mJy on April 7 (MJD~60407.93), before declining to $10.23 \pm 0.32$~mJy by April 26 (MJD~60426.76). Although the amplitudes of the secondary brightening differ between the observing frequencies, the consistent rise--decline pattern detected by two independently calibrated radio facilities indicates that the re-brightening is intrinsic to the source. This behavior is consistent with a second episode of enhanced jet activity, suggesting renewed coupling between the accretion flow and the outflow during the later stages of the outburst. The occurrence of this secondary radio flare, approximately one month after the initial outburst, indicates that the source likely experienced multiple episodes of jet activity rather than a single discrete ejection event. The secondary radio brightening is therefore consistent with renewed jet activity, similar to the multiple radio flares observed in several black hole X-ray binaries during state transitions \citep{Fender2004,Fender2009,Hughes2025}.

Although the \meerkat{} L-band images do not resolve any extended jet structure, together with its evolution and the contemporaneous {\it ATCA} flare, it is consistent with synchrotron emission from transient relativistic ejecta powered by the accretion flow \citep{Gallo2003,Fender2004}. The detection of relativistic reflection together with quasi-simultaneous radio emission is consistent with close coupling between the accretion flow and the outflow. In this context, one possible interpretation is that the compact corona is physically connected to the base of the jet, with inverse Compton and synchrotron emission from the jet base contributing to the hard X-ray continuum that irradiates the accretion disk and produces the observed reflection features \citep{Markoff2003,Markoff2004,Davidson2025}.

The quasi-simultaneous radio (\meerkat) and X-ray observations provide an opportunity to probe the jet activity during a certain accretion state. The source was detected in multiple \meerkat\ epochs, supplemented by {\it ATCA} and {\it ASKAP}/VAST observations, throughout the outburst, enabling us to investigate the evolution of its radio emission and its connection with the accretion flow. The quasi-simultaneous radio and X-ray measurements place \src\ on the radio/X-ray luminosity plane within the locus occupied by black hole X-ray binaries, consistent with the empirical $L_{\rm R}$--$L_{\rm X}$ correlation \citep{Corbel2003,Gallo2003}. This correlation is widely interpreted as evidence for a fundamental coupling between accretion and jet production in black hole systems, with the radio and X-ray luminosities tracing the jet power and the accretion rate, respectively. The location of \src\ on this relation, therefore, provides additional evidence supporting its classification as a black hole. Furthermore, the detection of radio emission during the soft/intermediate state indicates that jet activity had not yet been completely quenched, consistent with the evolving accretion--ejection coupling commonly observed during spectral state transitions in black hole X-ray binaries.
\vspace{5mm}
\software{{\sc HEASoft} V6.36 \citep{heasoft}, {\sc LaxpcSoft}, {\sc xspec} \citep[v12.10.0c;][]{Arnaud1996},
{\sc relxill}
\citep[v2.3.0;][]{gar14c,Da14},  NumPy and SciPy \citep{virtanen20}, Matplotlib \citep{hunter07}, IPython \citep{perez07}.}
\section*{Acknowledgements}
 We thank the referee for his/her valuable comments and suggestions, which improved the manuscript. The research work at the Physical Research Laboratory, Ahmedabad, is funded by the Department of Space, Government of India. MM acknowledges Ajay Ratheesh for his assistance with the \astrosat{} CZTI data reduction. This work is based on observations obtained with the AstroSat mission of the Indian Space Research Organisation (ISRO), archived at the Indian Space Science Data Center (ISSDC). We thank the \astrosat{} team for approving our ToO observation request and promptly scheduling the observations. This work has made use of the calibration databases and auxiliary analysis tools developed, maintained, and distributed by the \lxp{} team. We also acknowledge the use of data from the SXT, developed at TIFR, Mumbai, and thank the SXT POC at TIFR for verifying and releasing the data through the ISSDC archive and for providing the necessary analysis software. We acknowledge the use of data from the CZTI aboard \astrosat{} and thank the CZTI team for developing, maintaining, and distributing the calibration database and data analysis software. This research has made use of data obtained with \nustar{}, a project led by Caltech, funded by NASA, and managed by NASA/JPL, and has utilized the {\tt NUSTARDAS} software package, jointly developed by the ASDC (Italy) and Caltech (USA). 
We acknowledge the use of public data from the \nicer, and \nustar~ data archives. The \textit{MeerKAT} telescope is operated by the South African Radio Astronomy Observatory, which is a facility of the National Research Foundation, an agency of the Department of Science and Innovation. SM \& SP acknowledge the use of the Pegasus high-performance computing facility at IUCAA for this work.
%
\bibliographystyle{aasjournal}
\bibliography{my-references}

@ARTICLE{car24,
       author = {{Carotenuto}, F. and {Russell}, T.~D.},
        title = "{ATCA detection of an extremely bright radio flare from Swift J151857.0-572147}",
      journal = {The Astronomer's Telegram},
         year = 2024,
        month = mar,
       volume = {16518},
        pages = {1},
       adsurl = {https://ui.adsabs.harvard.edu/abs/2024ATel16518....1C}
}

@ARTICLE{pay24,
       author = {{Saikia}, Payaswini and {Russell}, D.~M. and {Baglio}, M.~C. and {Alabarta}, Kevin and {Rout}, S. and {Lewis}, F.},
        title = "{Search for the optical counterpart of the transient Swift J151857.0-572147}",
      journal = {The Astronomer's Telegram},
         year = 2024,
        month = mar,
       volume = {16516},
        pages = {1},
       adsurl = {https://ui.adsabs.harvard.edu/abs/2024ATel16516....1S}
}

@ARTICLE{Anumarlapudi24,
       author = {{Anumarlapudi}, Akash and {Kaplan}, David and {Sivakoff}, Gregory and {Dobie}, Dougal and {Qui}, Hao and {Murphy}, Tara and {Driessen}, Laura and {collaboration}, Emil Lenc VAST},
        title = "{ASKAP VAST radio detections of Swift J151857.0-572147}",
      journal = {The Astronomer's Telegram},
         year = 2024,
        month = may,
       volume = {16617},
        pages = {1},
       adsurl = {https://ui.adsabs.harvard.edu/abs/2024ATel16617....1A}
}

@ARTICLE{bag24,
       author = {{Baglio}, M.~C. and {D'Avanzo}, P. and {Ferro}, M. and {Campana}, S. and {Covino}, S. and {Fugazza}, D. and {Motta}, S. and {Mariani}, I. and {Messa}, M.~M. and {Russell}, D.~M. and {Saikia}, P. and {Alabarta}, K. and {Rout}, S.},
        title = "{The near-infrared counterpart of the newly discovered transient Swift J151857.0-572147}",
      journal = {The Astronomer's Telegram},
         year = 2024,
        month = mar,
       volume = {16506},
        pages = {1},
       adsurl = {https://ui.adsabs.harvard.edu/abs/2024ATel16506....1B}
}

@ARTICLE{cow24,
       author = {{Cowie}, F.~J. and {Carotenuto}, F. and {Fender}, R.~P. and {Heywood}, I. and {Hughes}, A.~K. and {Sivakoff}, G.~R. and {X-KAT Collaboration}},
        title = "{Discovery of the radio counterpart of Swift J151857.0-572147 with MeerKAT}",
      journal = {The Astronomer's Telegram},
         year = 2024,
        month = mar,
       volume = {16503},
        pages = {1},
       adsurl = {https://ui.adsabs.harvard.edu/abs/2024ATel16503....1C}
}

@ARTICLE{sha73,
   author = {{Shakura}, N.~I. and {Sunyaev}, R.~A.},
    title = "{Black holes in binary systems. Observational appearance.}",
  journal = {\aap},
     year = 1973,
   volume = 24,
    pages = {337-355},
   adsurl = {http://adsabs.harvard.edu/cgi-bin/nph-bib_query?bibcode=1973A%26A....24..337S&db_key=AST}
}

@ARTICLE{rem06,
   author = {{Remillard}, R.~A. and {McClintock}, J.~E.},
    title = "{X-Ray Properties of Black-Hole Binaries}",
  journal = {\araa},
   eprint = {arXiv:astro-ph/0606352},
     year = 2006,
    month = sep,
   volume = 44,
    pages = {49-92},
      doi = {10.1146/annurev.astro.44.051905.092532},
   adsurl = {http://cdsads.u-strasbg.fr/abs/2006ARA%26A..44...49R}
}

@INPROCEEDINGS{gar14c,
   author = {{Garc{\'{\i}}a}, J. and {McClintock}, J.~E. and {Steiner}, J.~F. and
	{Remillard}, R.~A. and {Grinberg}, V.},
    title = "{Modeling Reflection Signatures in the RXTE Spectra from X-ray Binaries: The GX 339-4 Case}",
booktitle = {AAS/High Energy Astrophysics Division},
     year = 2014,
   series = {AAS/High Energy Astrophysics Division},
   volume = 14,
    month = aug,
    pages = {205.02},
   adsurl = {http://adsabs.harvard.edu/abs/2014HEAD...1420502G}
}

@ARTICLE{ken24,
       author = {{Kennea}, J.~A. and {Lien}, A.~Y. and {D'Elia}, V. and {Melandri}, A. and {Page}, K.~L. and {Siegel}, M.~H.},
        title = "{Swift J151857.0-572147: Swift detection of a new galactic X-ray transient}",
      journal = {The Astronomer's Telegram},
         year = 2024,
        month = mar,
       volume = {16500},
        pages = {1},
       adsurl = {https://ui.adsabs.harvard.edu/abs/2024ATel16500....1K}
}

@ARTICLE{2016A&A...587A.151K,
       author = {{Kaastra}, J.~S. and {Bleeker}, J.~A.~M.},
        title = "{Optimal binning of X-ray spectra and response matrix design}",
      journal = {\aap},
         year = 2016,
        month = mar,
       volume = {587},
          eid = {A151},
        pages = {A151},
          doi = {10.1051/0004-6361/201527395},
archivePrefix = {arXiv},
       eprint = {1601.05309},
 primaryClass = {astro-ph.IM},
       adsurl = {https://ui.adsabs.harvard.edu/abs/2016A&A...587A.151K}
}

@ARTICLE{Bha17,
       author = {{Bhalerao}, V. and {Bhattacharya}, D. and {Vibhute}, A. and {Pawar}, P. and {Rao}, A.~R. and {Hingar}, M.~K. and {Khanna}, Rakesh and {Kutty}, A.~P.~K. and {Malkar}, J.~P. and {Patil}, M.~H. and {Arora}, Y.~K. and {Sinha}, S. and {Priya}, P. and {Samuel}, Essy and {Sreekumar}, S. and {Vinod}, P. and {Mithun}, N.~P.~S. and {Vadawale}, S.~V. and {Vagshette}, N. and {Navalgund}, K.~H. and {Sarma}, K.~S. and {Pandiyan}, R. and {Seetha}, S. and {Subbarao}, K.},
        title = "{The Cadmium Zinc Telluride Imager on AstroSat}",
      journal = {Journal of Astrophysics and Astronomy},
         year = 2017,
        month = jun,
       volume = {38},
       number = {2},
          eid = {31},
        pages = {31},
          doi = {10.1007/s12036-017-9447-8},
archivePrefix = {arXiv},
       eprint = {1608.03408},
 primaryClass = {astro-ph.IM},
       adsurl = {https://ui.adsabs.harvard.edu/abs/2017JApA...38...31B}
}

@INPROCEEDINGS{Singh2016,
       author = {{Singh}, Kulinder Pal and {Stewart}, Gordon C. and {Chandra}, Sunil and {Mukerjee}, Kallol and {Kotak}, Sanket and {Beardmore}, Andy P. and {Chitnis}, Varsha and {Dewangan}, Gulab C. and {Bhattacharyya}, Sudip and {Mirza}, Irfan and {Kamble}, Nilima and {Navalkar}, Vinita and {Shah}, Harshit and {Vishwakarma}, S. and {Koyande}, J.},
        title = "{In-orbit performance of SXT aboard AstroSat}",
    booktitle = {Space Telescopes and Instrumentation 2016: Ultraviolet to Gamma Ray},
         year = 2016,
       editor = {{den Herder}, Jan-Willem A. and {Takahashi}, Tadayuki and {Bautz}, Marshall},
       series = {Society of Photo-Optical Instrumentation Engineers (SPIE) Conference Series},
       volume = {9905},
        month = jul,
          eid = {99051E},
        pages = {99051E},
          doi = {10.1117/12.2235309},
       adsurl = {https://ui.adsabs.harvard.edu/abs/2016SPIE.9905E..1ES}
}

@ARTICLE{Singh2017,
       author = {{Singh}, K.~P. and {Stewart}, G.~C. and {Westergaard}, N.~J. and {Bhattacharayya}, S. and {Chandra}, S. and {Chitnis}, V.~R. and {Dewangan}, G.~C. and {Kothare}, A.~T. and {Mirza}, I.~M. and {Mukerjee}, K. and {Navalkar}, V. and {Shah}, H. and {Abbey}, A.~F. and {Beardmore}, A.~P. and {Kotak}, S. and {Kamble}, N. and {Vishwakarama}, S. and {Pathare}, D.~P. and {Risbud}, V.~M. and {Koyande}, J.~P. and {Stevenson}, T. and {Bicknell}, C. and {Crawford}, T. and {Hansford}, G. and {Peters}, G. and {Sykes}, J. and {Agarwal}, P. and {Sebastian}, M. and {Rajarajan}, A. and {Nagesh}, G. and {Narendra}, S. and {Ramesh}, M. and {Rai}, R. and {Navalgund}, K.~H. and {Sarma}, K.~S. and {Pandiyan}, R. and {Subbarao}, K. and {Gupta}, T. and {Thakkar}, N. and {Singh}, A.~K. and {Bajpai}, A.},
        title = "{Soft X-ray Focusing Telescope Aboard AstroSat: Design, Characteristics and Performance}",
      journal = {Journal of Astrophysics and Astronomy},
         year = 2017,
        month = jun,
       volume = {38},
       number = {2},
          eid = {29},
        pages = {29},
          doi = {10.1007/s12036-017-9448-7},
       adsurl = {https://ui.adsabs.harvard.edu/abs/2017JApA...38...29S}
}

@INPROCEEDINGS{Yadav2016,
       author = {{Yadav}, J.~S. and {Agrawal}, P.~C. and {Antia}, H.~M. and {Chauhan}, Jai Verdhan and {Dedhia}, Dhiraj and {Katoch}, Tilak and {Madhwani}, P. and {Manchanda}, R.~K. and {Misra}, Ranjeev and {Pahari}, Mayukh and {Paul}, B. and {Shah}, Parag},
        title = "{Large Area X-ray Proportional Counter (LAXPC) instrument onboard ASTROSAT}",
    booktitle = {Space Telescopes and Instrumentation 2016: Ultraviolet to Gamma Ray},
         year = 2016,
       editor = {{den Herder}, Jan-Willem A. and {Takahashi}, Tadayuki and {Bautz}, Marshall},
       series = {Society of Photo-Optical Instrumentation Engineers (SPIE) Conference Series},
       volume = {9905},
        month = jul,
          eid = {99051D},
        pages = {99051D},
          doi = {10.1117/12.2231857},
       adsurl = {https://ui.adsabs.harvard.edu/abs/2016SPIE.9905E..1DY}
}

@ARTICLE{Antia2017,
       author = {{Antia}, H.~M. and {Yadav}, J.~S. and {Agrawal}, P.~C. and {Verdhan Chauhan}, Jai and {Manchanda}, R.~K. and {Chitnis}, Varsha and {Paul}, Biswajit and {Dedhia}, Dhiraj and {Shah}, Parag and {Gujar}, V.~M. and {Katoch}, Tilak and {Kurhade}, V.~N. and {Madhwani}, Pankaj and {Manojkumar}, T.~K. and {Nikam}, V.~A. and {Pandya}, A.~S. and {Parmar}, J.~V. and {Pawar}, D.~M. and {Pahari}, Mayukh and {Misra}, Ranjeev and {Navalgund}, K.~H. and {Pandiyan}, R. and {Sharma}, K.~S. and {Subbarao}, K.},
        title = "{Calibration of the Large Area X-Ray Proportional Counter (LAXPC) Instrument on board AstroSat}",
      journal = {\apjs},
         year = 2017,
        month = jul,
       volume = {231},
       number = {1},
          eid = {10},
        pages = {10},
          doi = {10.3847/1538-4365/aa7a0e},
archivePrefix = {arXiv},
       eprint = {1702.08624},
 primaryClass = {astro-ph.IM},
       adsurl = {https://ui.adsabs.harvard.edu/abs/2017ApJS..231...10A}
}

@ARTICLE{Agrawal2006,
       author = {{Agrawal}, P.~C.},
        title = "{A broad spectral band Indian Astronomy satellite {\textquoteleft}Astrosat{\textquoteright}}",
      journal = {Advances in Space Research},
         year = 2006,
        month = jan,
       volume = {38},
       number = {12},
        pages = {2989-2994},
          doi = {10.1016/j.asr.2006.03.038},
       adsurl = {https://ui.adsabs.harvard.edu/abs/2006AdSpR..38.2989A}
}

@INPROCEEDINGS{Singh2014,
       author = {{Singh}, Kulinder Pal and {Tandon}, S.~N. and {Agrawal}, P.~C. and {Antia}, H.~M. and {Manchanda}, R.~K. and {Yadav}, J.~S. and {Seetha}, S. and {Ramadevi}, M.~C. and {Rao}, A.~R. and {Bhattacharya}, D. and {Paul}, B. and {Sreekumar}, P. and {Bhattacharyya}, S. and {Stewart}, G.~C. and {Hutchings}, J. and {Annapurni}, S.~A. and {Ghosh}, S.~K. and {Murthy}, J. and {Pati}, A. and {Rao}, N.~K. and {Stalin}, C.~S. and {Girish}, V. and {Sankarasubramanian}, K. and {Vadawale}, S. and {Bhalerao}, V.~B. and {Dewangan}, G.~C. and {Dedhia}, D.~K. and {Hingar}, M.~K. and {Katoch}, T.~B. and {Kothare}, A.~T. and {Mirza}, I. and {Mukerjee}, K. and {Shah}, H. and {Shah}, P. and {Mohan}, R. and {Sangal}, A.~K. and {Nagabhusana}, S. and {Sriram}, S. and {Malkar}, J.~P. and {Sreekumar}, S. and {Abbey}, A.~F. and {Hansford}, G.~M. and {Beardmore}, A.~P. and {Sharma}, M.~R. and {Murthy}, S. and {Kulkarni}, R. and {Meena}, G. and {Babu}, V.~C. and {Postma}, J.},
        title = "{ASTROSAT mission}",
    booktitle = {Space Telescopes and Instrumentation 2014: Ultraviolet to Gamma Ray},
         year = 2014,
       editor = {{Takahashi}, Tadayuki and {den Herder}, Jan-Willem A. and {Bautz}, Mark},
       series = {Society of Photo-Optical Instrumentation Engineers (SPIE) Conference Series},
       volume = {9144},
        month = jul,
          eid = {91441S},
        pages = {91441S},
          doi = {10.1117/12.2062667},
       adsurl = {https://ui.adsabs.harvard.edu/abs/2014SPIE.9144E..1SS}
}

@ARTICLE{Da14,
       author = {{Dauser}, T. and {Garcia}, J. and {Parker}, M.~L. and {Fabian}, A.~C. and {Wilms}, J.},
        title = "{The role of the reflection fraction in constraining black hole spin.}",
      journal = {\mnras},
         year = 2014,
        month = oct,
       volume = {444},
        pages = {L100-L104},
          doi = {10.1093/mnrasl/slu125},
archivePrefix = {arXiv},
       eprint = {1408.2347},
 primaryClass = {astro-ph.HE},
       adsurl = {https://ui.adsabs.harvard.edu/abs/2014MNRAS.444L.100D}
}

@ARTICLE{Ga14,
       author = {{Garc{\'\i}a}, J. and {Dauser}, T. and {Lohfink}, A. and {Kallman}, T.~R. and {Steiner}, J.~F. and {McClintock}, J.~E. and {Brenneman}, L. and {Wilms}, J. and {Eikmann}, W. and {Reynolds}, C.~S. and {Tombesi}, F.},
        title = "{Improved Reflection Models of Black Hole Accretion Disks: Treating the Angular Distribution of X-Rays}",
      journal = {\apj},
         year = 2014,
        month = feb,
       volume = {782},
       number = {2},
          eid = {76},
        pages = {76},
          doi = {10.1088/0004-637X/782/2/76},
archivePrefix = {arXiv},
       eprint = {1312.3231},
 primaryClass = {astro-ph.HE},
       adsurl = {https://ui.adsabs.harvard.edu/abs/2014ApJ...782...76G}
}

@ARTICLE{Good2010,
       author = {{Goodman}, Jonathan and {Weare}, Jonathan},
        title = "{Ensemble samplers with affine invariance}",
      journal = {Communications in Applied Mathematics and Computational Science},
         year = 2010,
        month = jan,
       volume = {5},
       number = {1},
        pages = {65-80},
          doi = {10.2140/camcos.2010.5.65},
       adsurl = {https://ui.adsabs.harvard.edu/abs/2010CAMCS...5...65G}
}

@ARTICLE{Ha13,
       author = {{Harrison}, Fiona A. and {Craig}, William W. and {Christensen}, Finn E. and {Hailey}, Charles J. and {Zhang}, William W. and {Boggs}, Steven E. and {Stern}, Daniel and {Cook}, W. Rick and {Forster}, Karl and {Giommi}, Paolo and {Grefenstette}, Brian W. and {Kim}, Yunjin and {Kitaguchi}, Takao and {Koglin}, Jason E. and {Madsen}, Kristin K. and {Mao}, Peter H. and {Miyasaka}, Hiromasa and {Mori}, Kaya and {Perri}, Matteo and {Pivovaroff}, Michael J. and {Puccetti}, Simonetta and {Rana}, Vikram R. and {Westergaard}, Niels J. and {Willis}, Jason and {Zoglauer}, Andreas and {An}, Hongjun and {Bachetti}, Matteo and {Barri{\`e}re}, Nicolas M. and {Bellm}, Eric C. and {Bhalerao}, Varun and {Brejnholt}, Nicolai F. and {Fuerst}, Felix and {Liebe}, Carl C. and {Markwardt}, Craig B. and {Nynka}, Melania and {Vogel}, Julia K. and {Walton}, Dominic J. and {Wik}, Daniel R. and {Alexander}, David M. and {Cominsky}, Lynn R. and {Hornschemeier}, Ann E. and {Hornstrup}, Allan and {Kaspi}, Victoria M. and {Madejski}, Greg M. and {Matt}, Giorgio and {Molendi}, Silvano and {Smith}, David M. and {Tomsick}, John A. and {Ajello}, Marco and {Ballantyne}, David R. and {Balokovi{\'c}}, Mislav and {Barret}, Didier and {Bauer}, Franz E. and {Blandford}, Roger D. and {Brandt}, W. Niel and {Brenneman}, Laura W. and {Chiang}, James and {Chakrabarty}, Deepto and {Chenevez}, Jerome and {Comastri}, Andrea and {Dufour}, Francois and {Elvis}, Martin and {Fabian}, Andrew C. and {Farrah}, Duncan and {Fryer}, Chris L. and {Gotthelf}, Eric V. and {Grindlay}, Jonathan E. and {Helfand}, David J. and {Krivonos}, Roman and {Meier}, David L. and {Miller}, Jon M. and {Natalucci}, Lorenzo and {Ogle}, Patrick and {Ofek}, Eran O. and {Ptak}, Andrew and {Reynolds}, Stephen P. and {Rigby}, Jane R. and {Tagliaferri}, Gianpiero and {Thorsett}, Stephen E. and {Treister}, Ezequiel and {Urry}, C. Megan},
        title = "{The Nuclear Spectroscopic Telescope Array (NuSTAR) High-energy X-Ray Mission}",
      journal = {\apj},
         year = 2013,
        month = jun,
       volume = {770},
       number = {2},
          eid = {103},
        pages = {103},
          doi = {10.1088/0004-637X/770/2/103},
archivePrefix = {arXiv},
       eprint = {1301.7307},
 primaryClass = {astro-ph.IM},
       adsurl = {https://ui.adsabs.harvard.edu/abs/2013ApJ...770..103H}
}

@ARTICLE{virtanen20,
  author  = {Virtanen, Pauli and Gommers, Ralf and Oliphant, Travis E. and
            Haberland, Matt and Reddy, Tyler and Cournapeau, David and
            Burovski, Evgeni and Peterson, Pearu and Weckesser, Warren and
            Bright, Jonathan and {van der Walt}, St{\'e}fan J. and
            Brett, Matthew and Wilson, Joshua and Millman, K. Jarrod and
            Mayorov, Nikolay and Nelson, Andrew R. J. and Jones, Eric and
            Kern, Robert and Larson, Eric and Carey, C J and
            Polat, {\.I}lhan and Feng, Yu and Moore, Eric W. and
            {VanderPlas}, Jake and Laxalde, Denis and Perktold, Josef and
            Cimrman, Robert and Henriksen, Ian and Quintero, E. A. and
            Harris, Charles R. and Archibald, Anne M. and
            Ribeiro, Ant{\^o}nio H. and Pedregosa, Fabian and
            {van Mulbregt}, Paul and {SciPy 1.0 Contributors}},
  title   = {{{SciPy} 1.0: Fundamental Algorithms for Scientific
            Computing in Python}},
  journal = {Nature Methods},
  year    = {2020},
  volume  = {17},
  pages   = {261--272},
  adsurl  = {https://rdcu.be/b08Wh},
  doi     = {10.1038/s41592-019-0686-2},
}

@ARTICLE{hunter07,
       author = {{Hunter}, John D.},
        title = "{Matplotlib: A 2D Graphics Environment}",
      journal = {Computing in Science and Engineering},
         year = 2007,
        month = may,
       volume = {9},
       number = {3},
        pages = {90-95},
          doi = {10.1109/MCSE.2007.55},
       adsurl = {https://ui.adsabs.harvard.edu/abs/2007CSE.....9...90H}
}

@ARTICLE{perez07,
       author = {{Perez}, Fernando and {Granger}, Brian E.},
        title = "{IPython: A System for Interactive Scientific Computing}",
      journal = {Computing in Science and Engineering},
         year = 2007,
        month = jan,
       volume = {9},
       number = {3},
        pages = {21-29},
          doi = {10.1109/MCSE.2007.53},
       adsurl = {https://ui.adsabs.harvard.edu/abs/2007CSE.....9c..21P}
}

@MISC{heasoft,
       author = {{NASA High Energy Astrophysics Science Archive Research Center (HEASARC)}},
        title = "{HEAsoft: Unified Release of FTOOLS and XANADU}",
 howpublished = {Astrophysics Source Code Library, record ascl:1408.004},
         year = 2014,
        month = aug,
          eid = {ascl:1408.004},
        pages = {ascl:1408.004},
archivePrefix = {ascl},
       eprint = {1408.004},
       adsurl = {https://ui.adsabs.harvard.edu/abs/2014ascl.soft08004N}
}

@ARTICLE{Re22,
       author = {{Remillard}, Ronald A. and {Loewenstein}, Michael and {Steiner}, James F. and {Prigozhin}, Gregory Y. and {LaMarr}, Beverly and {Enoto}, Teruaki and {Gendreau}, Keith C. and {Arzoumanian}, Zaven and {Markwardt}, Craig and {Basak}, Arkadip and {Stevens}, Abigail L. and {Ray}, Paul S. and {Altamirano}, Diego and {Buisson}, Douglas J.~K.},
        title = "{An Empirical Background Model for the NICER X-Ray Timing Instrument}",
      journal = {\aj},
         year = 2022,
        month = mar,
       volume = {163},
       number = {3},
          eid = {130},
        pages = {130},
          doi = {10.3847/1538-3881/ac4ae6},
archivePrefix = {arXiv},
       eprint = {2105.09901},
 primaryClass = {astro-ph.IM},
       adsurl = {https://ui.adsabs.harvard.edu/abs/2022AJ....163..130R}
}

@ARTICLE{Wilms2000,
       author = {{Wilms}, J. and {Allen}, A. and {McCray}, R.},
        title = "{On the Absorption of X-Rays in the Interstellar Medium}",
      journal = {\apj},
         year = 2000,
        month = oct,
       volume = {542},
       number = {2},
        pages = {914-924},
          doi = {10.1086/317016},
archivePrefix = {arXiv},
       eprint = {astro-ph/0008425},
 primaryClass = {astro-ph},
       adsurl = {https://ui.adsabs.harvard.edu/abs/2000ApJ...542..914W}
}

@ARTICLE{Zycki1999,
       author = {{{\.Z}ycki}, Piotr T. and {Done}, Chris and {Smith}, David A.},
        title = "{The 1989 May outburst of the soft X-ray transient GS 2023+338 (V404 Cyg)}",
      journal = {\mnras},
         year = 1999,
        month = nov,
       volume = {309},
       number = {3},
        pages = {561-575},
          doi = {10.1046/j.1365-8711.1999.02885.x},
archivePrefix = {arXiv},
       eprint = {astro-ph/9904304},
 primaryClass = {astro-ph},
       adsurl = {https://ui.adsabs.harvard.edu/abs/1999MNRAS.309..561Z}
}

@ARTICLE{Zdziarski1996,
       author = {{Zdziarski}, A.~A. and {Johnson}, W.~N. and {Magdziarz}, P.},
        title = "{Broad-band {\ensuremath{\gamma}}-ray and X-ray spectra of NGC 4151 and their implications for physical processes and geometry.}",
      journal = {\mnras},
         year = 1996,
        month = nov,
       volume = {283},
       number = {1},
        pages = {193-206},
          doi = {10.1093/mnras/283.1.193},
archivePrefix = {arXiv},
       eprint = {astro-ph/9607015},
 primaryClass = {astro-ph},
       adsurl = {https://ui.adsabs.harvard.edu/abs/1996MNRAS.283..193Z}
}

@ARTICLE{Mitsuda1984,
       author = {{Mitsuda}, K. and {Inoue}, H. and {Koyama}, K. and {Makishima}, K. and {Matsuoka}, M. and {Ogawara}, Y. and {Shibazaki}, N. and {Suzuki}, K. and {Tanaka}, Y. and {Hirano}, T.},
        title = "{Energy spectra of low-mass binary X-ray sources observed from Tenma.}",
      journal = {\pasj},
         year = 1984,
        month = jan,
       volume = {36},
        pages = {741-759},
       adsurl = {https://ui.adsabs.harvard.edu/abs/1984PASJ...36..741M}
}

@ARTICLE{Makishima1986,
       author = {{Makishima}, K. and {Maejima}, Y. and {Mitsuda}, K. and {Bradt}, H.~V. and {Remillard}, R.~A. and {Tuohy}, I.~R. and {Hoshi}, R. and {Nakagawa}, M.},
        title = "{Simultaneous X-Ray and Optical Observations of GX 339-4 in an X-Ray High State}",
      journal = {\apj},
         year = 1986,
        month = sep,
       volume = {308},
        pages = {635},
          doi = {10.1086/164534},
       adsurl = {https://ui.adsabs.harvard.edu/abs/1986ApJ...308..635M}
}

@ARTICLE{Fabian2000,
       author = {{Fabian}, A.~C. and {Iwasawa}, K. and {Reynolds}, C.~S. and {Young}, A.~J.},
        title = "{Broad Iron Lines in Active Galactic Nuclei}",
      journal = {\pasp},
         year = 2000,
        month = sep,
       volume = {112},
       number = {775},
        pages = {1145-1161},
          doi = {10.1086/316610},
archivePrefix = {arXiv},
       eprint = {astro-ph/0004366},
 primaryClass = {astro-ph},
       adsurl = {https://ui.adsabs.harvard.edu/abs/2000PASP..112.1145F}
}

@ARTICLE{Miller2007,
       author = {{Miller}, J.~M.},
        title = "{Relativistic X-Ray Lines from the Inner Accretion Disks Around Black Holes}",
      journal = {\araa},
         year = 2007,
        month = sep,
       volume = {45},
       number = {1},
        pages = {441-479},
          doi = {10.1146/annurev.astro.45.051806.110555},
archivePrefix = {arXiv},
       eprint = {0705.0540},
 primaryClass = {astro-ph},
       adsurl = {https://ui.adsabs.harvard.edu/abs/2007ARA&A..45..441M}
}

@INPROCEEDINGS{Arnaud1996,
       author = {{Arnaud}, K.~A.},
        title = "{XSPEC: The First Ten Years}",
    booktitle = {Astronomical Data Analysis Software and Systems V},
         year = 1996,
       editor = {{Jacoby}, George H. and {Barnes}, Jeannette},
       series = {Astronomical Society of the Pacific Conference Series},
       volume = {101},
        month = jan,
        pages = {17},
       adsurl = {https://ui.adsabs.harvard.edu/abs/1996ASPC..101...17A}
}

@ARTICLE{Peng2024,
       author = {{Peng}, Jing-Qiang and {Zhang}, Shu and {Shui}, Qing-Cang and {Chen}, Yu-Peng and {Zhang}, Shuang-Nan and {Kong}, Ling-Da and {Santangelo}, A. and {Yu}, Zhuo-Li and {Ji}, Long and {Wang}, Peng-Ju and {Chang}, Zhi and {Li}, Jian and {Li}, Zhao-sheng},
        title = "{Insight-HXMT, NICER, and NuSTAR Views to the Newly Discovered Black Hole X-Ray Binary Swift J151857.0─572147}",
      journal = {\apjl},
         year = 2024,
        month = sep,
       volume = {973},
       number = {1},
          eid = {L7},
        pages = {L7},
          doi = {10.3847/2041-8213/ad74ec},
archivePrefix = {arXiv},
       eprint = {2503.03093},
 primaryClass = {astro-ph.HE},
       adsurl = {https://ui.adsabs.harvard.edu/abs/2024ApJ...973L...7P}
}

@ARTICLE{Chatterjee2025,
       author = {{Chatterjee}, Kaushik and {Suribhatla}, S. Pujitha and {Mondal}, Santanu and {Singh}, Chandra B.},
        title = "{Interpreting the Spectrotemporal Properties of the Black Hole Candidate Swift J151857.0-572147 during Its First Outburst in 2024}",
      journal = {\apj},
         year = 2025,
        month = jul,
       volume = {987},
       number = {1},
          eid = {44},
        pages = {44},
          doi = {10.3847/1538-4357/add699},
archivePrefix = {arXiv},
       eprint = {2406.17629},
 primaryClass = {astro-ph.HE},
       adsurl = {https://ui.adsabs.harvard.edu/abs/2025ApJ...987...44C}
}

@ARTICLE{DelSanto2024,
       author = {{Del Santo}, Melania and {Russell}, Thomas D. and {Marino}, Alessio and {Motta}, Sara},
        title = "{Swift/XRT observation of Swift J151857.0-572147 indicates that the source likely made a transition to the soft state}",
      journal = {The Astronomer's Telegram},
         year = 2024,
        month = mar,
       volume = {16519},
        pages = {1},
       adsurl = {https://ui.adsabs.harvard.edu/abs/2024ATel16519....1D}
}

@ARTICLE{Homan2005,
       author = {{Homan}, Jeroen and {Belloni}, Tomaso},
        title = "{The Evolution of Black Hole States}",
      journal = {\apss},
         year = 2005,
        month = nov,
       volume = {300},
       number = {1-3},
        pages = {107-117},
          doi = {10.1007/s10509-005-1197-4},
archivePrefix = {arXiv},
       eprint = {astro-ph/0412597},
 primaryClass = {astro-ph},
       adsurl = {https://ui.adsabs.harvard.edu/abs/2005Ap&SS.300..107H}
}

@ARTICLE{Belloni2005,
       author = {{Belloni}, T. and {Homan}, J. and {Casella}, P. and {van der Klis}, M. and {Nespoli}, E. and {Lewin}, W.~H.~G. and {Miller}, J.~M. and {M{\'e}ndez}, M.},
        title = "{The evolution of the timing properties of the black-hole transient GX 339-4 during its 2002/2003 outburst}",
      journal = {\aap},
         year = 2005,
        month = sep,
       volume = {440},
       number = {1},
        pages = {207-222},
          doi = {10.1051/0004-6361:20042457},
archivePrefix = {arXiv},
       eprint = {astro-ph/0504577},
 primaryClass = {astro-ph},
       adsurl = {https://ui.adsabs.harvard.edu/abs/2005A&A...440..207B}
}

@ARTICLE{Homan2001,
       author = {{Homan}, Jeroen and {Wijnands}, Rudy and {van der Klis}, Michiel and {Belloni}, Tomaso and {van Paradijs}, Jan and {Klein-Wolt}, Marc and {Fender}, Rob and {M{\'e}ndez}, Mariano},
        title = "{Correlated X-Ray Spectral and Timing Behavior of the Black Hole Candidate XTE J1550-564: A New Interpretation of Black Hole States}",
      journal = {\apjs},
         year = 2001,
        month = feb,
       volume = {132},
       number = {2},
        pages = {377-402},
          doi = {10.1086/318954},
archivePrefix = {arXiv},
       eprint = {astro-ph/0001163},
 primaryClass = {astro-ph},
       adsurl = {https://ui.adsabs.harvard.edu/abs/2001ApJS..132..377H}
}

@ARTICLE{Corbel2003,
       author = {{Corbel}, S. and {Nowak}, M.~A. and {Fender}, R.~P. and {Tzioumis}, A.~K. and {Markoff}, S.},
        title = "{Radio/X-ray correlation in the low/hard state of GX 339-4}",
      journal = {\aap},
         year = 2003,
        month = mar,
       volume = {400},
        pages = {1007-1012},
          doi = {10.1051/0004-6361:20030090},
archivePrefix = {arXiv},
       eprint = {astro-ph/0301436},
 primaryClass = {astro-ph},
       adsurl = {https://ui.adsabs.harvard.edu/abs/2003A&A...400.1007C}
}

@ARTICLE{Antia2021,
       author = {{Antia}, H.~M. and {Agrawal}, P.~C. and {Dedhia}, Dhiraj and {Katoch}, Tilak and {Manchanda}, R.~K. and {Misra}, Ranjeev and {Mukerjee}, Kallol and {Pahari}, Mayukh and {Roy}, Jayashree and {Shah}, P. and {Yadav}, J.~S.},
        title = "{Large Area X-ray Proportional Counter (LAXPC) in orbit performance: Calibration, background, analysis software}",
      journal = {Journal of Astrophysics and Astronomy},
         year = 2021,
        month = oct,
       volume = {42},
       number = {2},
          eid = {32},
        pages = {32},
          doi = {10.1007/s12036-021-09712-8},
archivePrefix = {arXiv},
       eprint = {2101.07514},
 primaryClass = {astro-ph.IM},
       adsurl = {https://ui.adsabs.harvard.edu/abs/2021JApA...42...32A}
}

@ARTICLE{Beri2021,
       author = {{Beri}, Aru and {Naik}, Sachindra and {Singh}, Kulinder Pal and {Jaisawal}, Gaurava K. and {Bhattacharyya}, Sudip and {Charles}, Philip and {Ho}, Wynn C.~G. and {Maitra}, Chandreyee and {Bhattacharya}, Dipankar and {Dewangan}, Gulab C. and {Middleton}, Matthew and {Altamirano}, Diego and {Gandhi}, Poshak and {Raichur}, Harsha},
        title = "{AstroSat observations of the first Galactic ULX pulsar Swift J0243.6+6124}",
      journal = {\mnras},
         year = 2021,
        month = jan,
       volume = {500},
       number = {1},
        pages = {565-575},
          doi = {10.1093/mnras/staa3254},
archivePrefix = {arXiv},
       eprint = {2010.08334},
 primaryClass = {astro-ph.HE},
       adsurl = {https://ui.adsabs.harvard.edu/abs/2021MNRAS.500..565B}
}

@ARTICLE{Sharma2024,
       author = {{Sharma}, Rahul and {Mandal}, Manoj and {Pal}, Sabyasachi and {Paul}, Biswajit and {Jaisawal}, G.~K. and {Ratheesh}, Ajay},
        title = "{Probing the energy and luminosity-dependent spectro-timing properties of RX J0440.9+4431 with AstroSat}",
      journal = {\mnras},
         year = 2024,
        month = oct,
       volume = {534},
       number = {2},
        pages = {1028-1042},
          doi = {10.1093/mnras/stae2175},
archivePrefix = {arXiv},
       eprint = {2409.11121},
 primaryClass = {astro-ph.HE},
       adsurl = {https://ui.adsabs.harvard.edu/abs/2024MNRAS.534.1028S}
}

@ARTICLE{Steiner2024,
       author = {{Steiner}, James F. and {Nathan}, Edward and {Hu}, Kun and {Krawczynski}, Henric and {Dov{\v{c}}iak}, Michal and {Veledina}, Alexandra and {Muleri}, Fabio and {Svoboda}, Jiri and {Alabarta}, Kevin and {Parra}, Maxime and {Bhargava}, Yash and {Matt}, Giorgio and {Poutanen}, Juri and {Petrucci}, Pierre-Olivier and {Tennant}, Allyn F. and {Baglio}, M. Cristina and {Baldini}, Luca and {Barnier}, Samuel and {Bhattacharyya}, Sudip and {Bianchi}, Stefano and {Brigitte}, Maimouna and {Cabezas}, Mauricio and {Cangemi}, Floriane and {Capitanio}, Fiamma and {Casey}, Jacob and {Rodriguez Cavero}, Nicole and {Castellano}, Simone and {Cavazzuti}, Elisabetta and {Chun}, Sohee and {Churazov}, Eugene and {Costa}, Enrico and {Di Lalla}, Niccol{\`o} and {Di Marco}, Alessandro and {Egron}, Elise and {Ewing}, Melissa and {Fabiani}, Sergio and {Garc{\'\i}a}, Javier A. and {Green}, David A. and {Grinberg}, Victoria and {Hadrava}, Petr and {Ingram}, Adam and {Kaaret}, Philip and {Kislat}, Fabian and {Kitaguchi}, Takao and {Kravtsov}, Vadim and {Kub{\'a}tov{\'a}}, Brankica and {La Monaca}, Fabio and {Latronico}, Luca and {Loktev}, Vladislav and {Malacaria}, Christian and {Marin}, Fr{\'e}d{\'e}ric and {Marinucci}, Andrea and {Maryeva}, Olga and {Mastroserio}, Guglielmo and {Mizuno}, Tsunefumi and {Negro}, Michela and {Omodei}, Nicola and {Podgorn{\'y}}, Jakub and {Rankin}, John and {Ratheesh}, Ajay and {Rhodes}, Lauren and {Russell}, David M. and {{\v{S}}lechta}, Miroslav and {Soffitta}, Paolo and {Spooner}, Sean and {Suleimanov}, Valery and {Tombesi}, Francesco and {Trushkin}, Sergei A. and {Weisskopf}, Martin C. and {Zane}, Silvia and {Zdziarski}, Andrzej A. and {Zhang}, Sixuan and {Zhang}, Wenda and {Zhou}, Menglei and {Agudo}, Iv{\'a}n and {Antonelli}, Lucio A. and {Bachetti}, Matteo and {Baumgartner}, Wayne H. and {Bellazzini}, Ronaldo and {Bongiorno}, Stephen D. and {Bonino}, Raffaella and {Brez}, Alessandro and {Bucciantini}, Niccol{\`o} and {Chen}, Chien-Ting and {Ciprini}, Stefano and {De Rosa}, Alessandra and {Del Monte}, Ettore and {Di Gesu}, Laura and {Donnarumma}, Immacolata and {Doroshenko}, Victor and {Ehlert}, Steven R. and {Enoto}, Teruaki and {Evangelista}, Yuri and {Ferrazzoli}, Riccardo and {Gunji}, Shuichi and {Hayashida}, Kiyoshi and {Heyl}, Jeremy and {Iwakiri}, Wataru and {Jorstad}, Svetlana G. and {Karas}, Vladimir and {Kolodziejczak}, Jeffery J. and {Liodakis}, Ioannis and {Maldera}, Simone and {Manfreda}, Alberto and {Marscher}, Alan P. and {Marshall}, Herman L. and {Massaro}, Francesco and {Mitsuishi}, Ikuyuki and {Ng}, Chi-Yung and {O'Dell}, Stephen L. and {Oppedisano}, Chiara and {Papitto}, Alessandro and {Pavlov}, George G. and {Peirson}, Abel L. and {Perri}, Matteo and {Pesce-Rollins}, Melissa and {Pilia}, Maura and {Possenti}, Andrea and {Puccetti}, Simonetta and {Ramsey}, Brian D. and {Roberts}, Oliver J. and {Romani}, Roger W. and {Sgr{\`o}}, Carmelo and {Slane}, Patrick and {Spandre}, Gloria and {Swartz}, Douglas A. and {Tamagawa}, Toru and {Tavecchio}, Fabrizio and {Taverna}, Roberto and {Tawara}, Yuzuru and {Thomas}, Nicholas E. and {Trois}, Alessio and {Tsygankov}, Sergey S. and {Turolla}, Roberto and {Vink}, Jacco and {Wu}, Kinwah and {Xie}, Fei},
        title = "{An IXPE-led X-Ray Spectropolarimetric Campaign on the Soft State of Cygnus X-1: X-Ray Polarimetric Evidence for Strong Gravitational Lensing}",
      journal = {\apjl},
         year = 2024,
        month = jul,
       volume = {969},
       number = {2},
          eid = {L30},
        pages = {L30},
          doi = {10.3847/2041-8213/ad58e4},
archivePrefix = {arXiv},
       eprint = {2406.12014},
 primaryClass = {astro-ph.HE},
       adsurl = {https://ui.adsabs.harvard.edu/abs/2024ApJ...969L..30S}
}

@ARTICLE{Mandal2023,
       author = {{Mandal}, Manoj and {Sharma}, Rahul and {Pal}, Sabyasachi and {Jaisawal}, G.~K. and {Gendreau}, Keith C. and {Ng}, Mason and {Sanna}, Andrea and {Malacaria}, Christian and {Tombesi}, Francesco and {Ferrara}, E.~C. and {Markwardt}, Craig B. and {Wolff}, Michael T. and {Coley}, Joel B.},
        title = "{Probing spectral and timing properties of the X-ray pulsar RX J0440.9 + 4431 in the giant outburst of 2022-2023}",
      journal = {\mnras},
         year = 2023,
        month = nov,
       volume = {526},
       number = {1},
        pages = {771-781},
          doi = {10.1093/mnras/stad2767},
archivePrefix = {arXiv},
       eprint = {2306.08083},
 primaryClass = {astro-ph.HE},
       adsurl = {https://ui.adsabs.harvard.edu/abs/2023MNRAS.526..771M}
}

@ARTICLE{Chhotaray2023,
       author = {{Chhotaray}, Birendra and {Jaisawal}, Gaurava K. and {Kumari}, Neeraj and {Naik}, Sachindra and {Kumar}, Vipin and {Jana}, Arghajit},
        title = "{Optical and X-ray studies of Be/X-ray binary 1A 0535+262 during its 2020 giant outburst}",
      journal = {\mnras},
         year = 2023,
        month = feb,
       volume = {518},
       number = {4},
        pages = {5089-5105},
          doi = {10.1093/mnras/stac3354},
archivePrefix = {arXiv},
       eprint = {2211.09491},
 primaryClass = {astro-ph.HE},
       adsurl = {https://ui.adsabs.harvard.edu/abs/2023MNRAS.518.5089C}
}

@software{Bahramian2022,
  author       = {Arash Bahramian and
                  Anthony Rushton},
  title        = {bersavosh/XRB-LrLx\_pub: update 20220908},
  month        = sep,
  year         = 2022,
  publisher    = {Zenodo},
  version      = {v220908},
  doi          = {10.5281/zenodo.7059313},
  url          = {https://doi.org/10.5281/zenodo.7059313}
}

@ARTICLE{Gallo2006,
       author = {{Gallo}, E. and {Fender}, R.~P. and {Miller-Jones}, J.~C.~A. and {Merloni}, A. and {Jonker}, P.~G. and {Heinz}, S. and {Maccarone}, T.~J. and {van der Klis}, M.},
        title = "{A radio-emitting outflow in the quiescent state of A0620-00: implications for modelling low-luminosity black hole binaries}",
      journal = {\mnras},
         year = 2006,
        month = aug,
       volume = {370},
       number = {3},
        pages = {1351-1360},
          doi = {10.1111/j.1365-2966.2006.10560.x},
archivePrefix = {arXiv},
       eprint = {astro-ph/0605376},
 primaryClass = {astro-ph},
       adsurl = {https://ui.adsabs.harvard.edu/abs/2006MNRAS.370.1351G}
}

@INCOLLECTION{Kalemci2022,
       author = {{Kalemci}, Emrah and {Kara}, Erin and {Tomsick}, John A.},
        title = "{Black Holes: Timing and Spectral Properties and Evolution}",
    booktitle = {Handbook of X-ray and Gamma-ray Astrophysics},
         year = 2022,
       editor = {{Bambi}, Cosimo and {Sangangelo}, Andrea},
          eid = {9},
        pages = {9},
          doi = {10.1007/978-981-16-4544-0_100-1},
       adsurl = {https://ui.adsabs.harvard.edu/abs/2022hxga.book....9K}
}

@ARTICLE{Motta2011,
       author = {{Motta}, S. and {Mu{\~n}oz-Darias}, T. and {Casella}, P. and {Belloni}, T. and {Homan}, J.},
        title = "{Low-frequency oscillations in black holes: a spectral-timing approach to the case of GX 339-4}",
      journal = {\mnras},
         year = 2011,
        month = dec,
       volume = {418},
       number = {4},
        pages = {2292-2307},
          doi = {10.1111/j.1365-2966.2011.19566.x},
archivePrefix = {arXiv},
       eprint = {1108.0540},
 primaryClass = {astro-ph.HE},
       adsurl = {https://ui.adsabs.harvard.edu/abs/2011MNRAS.418.2292M}
}

@ARTICLE{Casella2004,
       author = {{Casella}, P. and {Belloni}, T. and {Homan}, J. and {Stella}, L.},
        title = "{A study of the low-frequency quasi-periodic oscillations in the X-ray light curves of the black hole candidate <ASTROBJ>XTE J1859+226</ASTROBJ>}",
      journal = {\aap},
         year = 2004,
        month = nov,
       volume = {426},
        pages = {587-600},
          doi = {10.1051/0004-6361:20041231},
archivePrefix = {arXiv},
       eprint = {astro-ph/0407262},
 primaryClass = {astro-ph},
       adsurl = {https://ui.adsabs.harvard.edu/abs/2004A&A...426..587C}
}

@ARTICLE{Casella2005,
       author = {{Casella}, P. and {Belloni}, T. and {Stella}, L.},
        title = "{The ABC of Low-Frequency Quasi-periodic Oscillations in Black Hole Candidates: Analogies with Z Sources}",
      journal = {\apj},
         year = 2005,
        month = aug,
       volume = {629},
       number = {1},
        pages = {403-407},
          doi = {10.1086/431174},
archivePrefix = {arXiv},
       eprint = {astro-ph/0504318},
 primaryClass = {astro-ph},
       adsurl = {https://ui.adsabs.harvard.edu/abs/2005ApJ...629..403C}
}

@ARTICLE{2017MNRAS.471..301O,
       author = {{Offringa}, A.~R. and {Smirnov}, O.},
        title = "{An optimized algorithm for multiscale wideband deconvolution of radio astronomical images}",
      journal = {\mnras},
         year = 2017,
        month = oct,
       volume = {471},
       number = {1},
        pages = {301-316},
          doi = {10.1093/mnras/stx1547},
archivePrefix = {arXiv},
       eprint = {1706.06786},
 primaryClass = {astro-ph.IM},
       adsurl = {https://ui.adsabs.harvard.edu/abs/2017MNRAS.471..301O}
}

@INPROCEEDINGS{2022ASPC..532..541H,
       author = {{Hugo}, Benjamin V. and {Perkins}, S. and {Merry}, B. and {Mauch}, T. and {Smirnov}, O.~M.},
        title = "{Tricolour: An Optimized SumThreshold Flagger for MeerKAT}",
    booktitle = {Astronomical Data Analysis Software and Systems XXX},
         year = 2022,
       editor = {{Ruiz}, Jose Enrique and {Pierfedereci}, Francesco and {Teuben}, Peter},
       series = {Astronomical Society of the Pacific Conference Series},
       volume = {532},
        month = jul,
        pages = {541},
          doi = {10.48550/arXiv.2206.09179},
archivePrefix = {arXiv},
       eprint = {2206.09179},
 primaryClass = {astro-ph.IM},
       adsurl = {https://ui.adsabs.harvard.edu/abs/2022ASPC..532..541H}
}

@INPROCEEDINGS{2007ASPC..376..127M,
       author = {{McMullin}, J.~P. and {Waters}, B. and {Schiebel}, D. and {Young}, W. and {Golap}, K.},
        title = "{CASA Architecture and Applications}",
    booktitle = {Astronomical Data Analysis Software and Systems XVI},
         year = 2007,
       editor = {{Shaw}, R.~A. and {Hill}, F. and {Bell}, D.~J.},
       series = {Astronomical Society of the Pacific Conference Series},
       volume = {376},
        month = oct,
        pages = {127},
       adsurl = {https://ui.adsabs.harvard.edu/abs/2007ASPC..376..127M}
}

@ARTICLE{Kenyon2018,
       author = {{Kenyon}, J.~S. and {Smirnov}, O.~M. and {Grobler}, T.~L. and {Perkins}, S.~J.},
        title = "{CUBICAL - fast radio interferometric calibration suite exploiting complex optimization}",
      journal = {\mnras},
         year = 2018,
        month = aug,
       volume = {478},
       number = {2},
        pages = {2399-2415},
          doi = {10.1093/mnras/sty1221},
archivePrefix = {arXiv},
       eprint = {1805.03410},
 primaryClass = {astro-ph.IM},
       adsurl = {https://ui.adsabs.harvard.edu/abs/2018MNRAS.478.2399K}
}

@ARTICLE{2014MNRAS.444..606O,
       author = {{Offringa}, A.~R. and {McKinley}, B. and {Hurley-Walker}, N. and {Briggs}, F.~H. and {Wayth}, R.~B. and {Kaplan}, D.~L. and {Bell}, M.~E. and {Feng}, L. and {Neben}, A.~R. and {Hughes}, J.~D. and {Rhee}, J. and {Murphy}, T. and {Bhat}, N.~D.~R. and {Bernardi}, G. and {Bowman}, J.~D. and {Cappallo}, R.~J. and {Corey}, B.~E. and {Deshpande}, A.~A. and {Emrich}, D. and {Ewall-Wice}, A. and {Gaensler}, B.~M. and {Goeke}, R. and {Greenhill}, L.~J. and {Hazelton}, B.~J. and {Hindson}, L. and {Johnston-Hollitt}, M. and {Jacobs}, D.~C. and {Kasper}, J.~C. and {Kratzenberg}, E. and {Lenc}, E. and {Lonsdale}, C.~J. and {Lynch}, M.~J. and {McWhirter}, S.~R. and {Mitchell}, D.~A. and {Morales}, M.~F. and {Morgan}, E. and {Kudryavtseva}, N. and {Oberoi}, D. and {Ord}, S.~M. and {Pindor}, B. and {Procopio}, P. and {Prabu}, T. and {Riding}, J. and {Roshi}, D.~A. and {Shankar}, N. Udaya and {Srivani}, K.~S. and {Subrahmanyan}, R. and {Tingay}, S.~J. and {Waterson}, M. and {Webster}, R.~L. and {Whitney}, A.~R. and {Williams}, A. and {Williams}, C.~L.},
        title = "{WSCLEAN: an implementation of a fast, generic wide-field imager for radio astronomy}",
      journal = {\mnras},
         year = 2014,
        month = oct,
       volume = {444},
       number = {1},
        pages = {606-619},
          doi = {10.1093/mnras/stu1368},
archivePrefix = {arXiv},
       eprint = {1407.1943},
 primaryClass = {astro-ph.IM},
       adsurl = {https://ui.adsabs.harvard.edu/abs/2014MNRAS.444..606O}
}

@ARTICLE{Gallo2003,
       author = {{Gallo}, E. and {Fender}, R.~P. and {Pooley}, G.~G.},
        title = "{A universal radio-X-ray correlation in low/hard state black hole binaries}",
      journal = {\mnras},
         year = 2003,
        month = sep,
       volume = {344},
       number = {1},
        pages = {60-72},
          doi = {10.1046/j.1365-8711.2003.06791.x},
archivePrefix = {arXiv},
       eprint = {astro-ph/0305231},
 primaryClass = {astro-ph},
       adsurl = {https://ui.adsabs.harvard.edu/abs/2003MNRAS.344...60G}
}

@ARTICLE{Fender2004,
       author = {{Fender}, R.~P. and {Belloni}, T.~M. and {Gallo}, E.},
        title = "{Towards a unified model for black hole X-ray binary jets}",
      journal = {\mnras},
         year = 2004,
        month = dec,
       volume = {355},
       number = {4},
        pages = {1105-1118},
          doi = {10.1111/j.1365-2966.2004.08384.x},
archivePrefix = {arXiv},
       eprint = {astro-ph/0409360},
 primaryClass = {astro-ph},
       adsurl = {https://ui.adsabs.harvard.edu/abs/2004MNRAS.355.1105F}
}

@ARTICLE{Davidson2025,
       author = {{Davidson}, Eric M. and {Chauhan}, Jaiverdhan and {Lohfink}, Anne and {Russell}, Thomas D. and {Starling}, Rhaana and {Johnson}, Charlotte},
        title = "{Establishing a Connection between the Jet and the Corona in Black Hole Low-mass X-Ray Binaries}",
      journal = {\apj},
         year = 2025,
        month = nov,
       volume = {994},
       number = {1},
          eid = {54},
        pages = {54},
          doi = {10.3847/1538-4357/ae0a4d},
archivePrefix = {arXiv},
       eprint = {2505.07962},
 primaryClass = {astro-ph.HE},
       adsurl = {https://ui.adsabs.harvard.edu/abs/2025ApJ...994...54D}
}

@ARTICLE{Markoff2004,
       author = {{Markoff}, Sera and {Nowak}, Michael A.},
        title = "{Constraining X-Ray Binary Jet Models via Reflection}",
      journal = {\apj},
         year = 2004,
        month = jul,
       volume = {609},
       number = {2},
        pages = {972-976},
          doi = {10.1086/421099},
archivePrefix = {arXiv},
       eprint = {astro-ph/0403468},
 primaryClass = {astro-ph},
       adsurl = {https://ui.adsabs.harvard.edu/abs/2004ApJ...609..972M}
}

@ARTICLE{Markoff2003,
       author = {{Markoff}, Sera and {Nowak}, Michael and {Corbel}, St{\'e}phane and {Fender}, Rob and {Falcke}, Heino},
        title = "{Modeling the X-ray contribution of XRB jets}",
      journal = {\nar},
         year = 2003,
        month = oct,
       volume = {47},
       number = {6-7},
        pages = {491-493},
          doi = {10.1016/S1387-6473(03)00078-2},
archivePrefix = {arXiv},
       eprint = {astro-ph/0301479},
 primaryClass = {astro-ph},
       adsurl = {https://ui.adsabs.harvard.edu/abs/2003NewAR..47..491M}
}

@ARTICLE{Fender2009,
       author = {{Fender}, R.~P. and {Homan}, J. and {Belloni}, T.~M.},
        title = "{Jets from black hole X-ray binaries: testing, refining and extending empirical models for the coupling to X-rays}",
      journal = {\mnras},
         year = 2009,
        month = jul,
       volume = {396},
       number = {3},
        pages = {1370-1382},
          doi = {10.1111/j.1365-2966.2009.14841.x},
archivePrefix = {arXiv},
       eprint = {0903.5166},
 primaryClass = {astro-ph.HE},
       adsurl = {https://ui.adsabs.harvard.edu/abs/2009MNRAS.396.1370F}
}

@ARTICLE{Hughes2025,
       author = {{Hughes}, Andrew K. and {Carotenuto}, Francesco and {Russell}, Thomas D. and {Tetarenko}, Alexandra J. and {Miller-Jones}, James C.~A. and {Bahramian}, Arash and {Bright}, Joe S. and {Cowie}, Fraser J. and {Fender}, Rob and {Gurwell}, Mark A. and {Khaulsay}, Jasvinderjit K. and {Kirby}, Anastasia and {Jones}, Serena and {Lescure}, Elodie and {McCollough}, Michael and {Plotkin}, Richard M. and {Rao}, Ramprasad and {Vrtilek}, Saeqa D. and {Williams-Baldwin}, David R.~A. and {Wood}, Callan M. and {Sivakoff}, Gregory R. and {Altamirano}, Diego and {Casella}, Piergiorgio and {Corbel}, St{\'e}phane and {DeBoer}, David R. and {Del Santo}, Melania and {Echibur{\'u}-Trujillo}, Constanza and {Farah}, Wael and {Gandhi}, Poshak and {Koljonen}, Karri I.~I. and {Maccarone}, Thomas and {Matthews}, James H. and {Markoff}, Sera B. and {Pollak}, Alexander W. and {Russell}, David M. and {Saikia}, Payaswini and {Castro Segura}, Noel and {Shaw}, Aarran W. and {Siemion}, Andrew and {Soria}, Roberto and {Tomsick}, John A. and {van den Eijnden}, Jakob},
        title = "{Comprehensive Radio Monitoring of the Black Hole X-Ray Binary Swift J1727.8{\ensuremath{-}}1613 during Its 2023─2024 Outburst}",
      journal = {\apj},
         year = 2025,
        month = jul,
       volume = {988},
       number = {1},
          eid = {109},
        pages = {109},
          doi = {10.3847/1538-4357/ade2e6},
archivePrefix = {arXiv},
       eprint = {2506.07798},
 primaryClass = {astro-ph.HE},
       adsurl = {https://ui.adsabs.harvard.edu/abs/2025ApJ...988..109H}
}

@ARTICLE{Furst2015,
       author = {{F{\"u}rst}, F. and {Nowak}, M.~A. and {Tomsick}, J.~A. and {Miller}, J.~M. and {Corbel}, S. and {Bachetti}, M. and {Boggs}, S.~E. and {Christensen}, F.~E. and {Craig}, W.~W. and {Fabian}, A.~C. and {Gandhi}, P. and {Grinberg}, V. and {Hailey}, C.~J. and {Harrison}, F.~A. and {Kara}, E. and {Kennea}, J.~A. and {Madsen}, K.~K. and {Pottschmidt}, K. and {Stern}, D. and {Walton}, D.~J. and {Wilms}, J. and {Zhang}, W.~W.},
        title = "{The Complex Accretion Geometry of GX 339-4 as Seen by NuSTAR and Swift}",
      journal = {\apj},
         year = 2015,
        month = aug,
       volume = {808},
       number = {2},
          eid = {122},
        pages = {122},
          doi = {10.1088/0004-637X/808/2/122},
archivePrefix = {arXiv},
       eprint = {1506.01381},
 primaryClass = {astro-ph.HE},
       adsurl = {https://ui.adsabs.harvard.edu/abs/2015ApJ...808..122F}
}

@ARTICLE{You2021,
       author = {{You}, Bei and {Tuo}, Yuoli and {Li}, Chengzhe and {Wang}, Wei and {Zhang}, Shuang-Nan and {Zhang}, Shu and {Ge}, Mingyu and {Luo}, Chong and {Liu}, Bifang and {Yuan}, Weimin and {Dai}, Zigao and {Liu}, Jifeng and {Qiao}, Erlin and {Jin}, Chichuan and {Liu}, Zhu and {Czerny}, Bozena and {Wu}, Qingwen and {Bu}, Qingcui and {Cai}, Ce and {Cao}, Xuelei and {Chang}, Zhi and {Chen}, Gang and {Chen}, Li and {Chen}, Tianxiang and {Chen}, Yibao and {Chen}, Yong and {Chen}, Yupeng and {Cui}, Wei and {Cui}, Weiwei and {Deng}, Jingkang and {Dong}, Yongwei and {Du}, Yuanyuan and {Fu}, Minxue and {Gao}, Guanhua and {Gao}, He and {Gao}, Min and {Gu}, Yudong and {Guan}, Ju and {Guo}, Chengcheng and {Han}, Dawei and {Huang}, Yue and {Huo}, Jia and {Jia}, Shumei and {Jiang}, Luhua and {Jiang}, Weichun and {Jin}, Jing and {Jin}, Yongjie and {Kong}, Lingda and {Li}, Bing and {Li}, Chengkui and {Li}, Gang and {Li}, Maoshun and {Li}, Tipei and {Li}, Wei and {Li}, Xian and {Li}, Xiaobo and {Li}, Xufang and {Li}, Yanguo and {Li}, Zhengwei and {Liang}, Xiaohua and {Liao}, Jinyuan and {Liu}, Congzhan and {Liu}, Guoqing and {Liu}, Hongwei and {Liu}, Xiaojing and {Liu}, Yinong and {Lu}, Bo and {Lu}, Fangjun and {Lu}, Xuefeng and {Luo}, Qi and {Luo}, Tao and {Ma}, Xiang and {Meng}, Bin and {Nang}, Yi and {Nie}, Jianyin and {Ou}, Ge and {Qu}, Jinlu and {Sai}, Na and {Shang}, Rencheng and {Song}, Liming and {Song}, Xinying and {Sun}, Liang and {Tan}, Ying and {Tao}, Lian and {Wang}, Chen and {Wang}, Guofeng and {Wang}, Juan and {Wang}, Lingjun and {Wang}, Wenshuai and {Wang}, Yusa and {Wen}, Xiangyang and {Wu}, Baiyang and {Wu}, Bobing and {Wu}, Mei and {Xiao}, Guangcheng and {Xiao}, Shuo and {Xiong}, Shaolin and {Xu}, Yupeng and {Yang}, Jiawei and {Yang}, Sheng and {Yang}, Yanji and {Yi}, Qibin and {Yin}, Qianqing and {You}, Yuan and {Zhang}, Aimei and {Zhang}, Chengmo and {Zhang}, Fan and {Zhang}, Hongmei and {Zhang}, Juan and {Zhang}, Tong and {Zhang}, Wanchang and {Zhang}, Wei and {Zhang}, Wenzhao and {Zhang}, Yi and {Zhang}, Yifei and {Zhang}, Yongjie and {Zhang}, Yue and {Zhang}, Zhao and {Zhang}, Ziliang and {Zhao}, Haisheng and {Zhao}, Xiaofan and {Zheng}, Shijie and {Zhou}, Dengke and {Zhou}, Jianfeng and {Zhu}, Yuxuan and {Zhu}, Yue},
        title = "{Insight-HXMT observations of jet-like corona in a black hole X-ray binary MAXI J1820+070}",
      journal = {Nature Communications},
         year = 2021,
        month = jan,
       volume = {12},
          eid = {1025},
        pages = {1025},
          doi = {10.1038/s41467-021-21169-5},
archivePrefix = {arXiv},
       eprint = {2102.07602},
 primaryClass = {astro-ph.HE},
       adsurl = {https://ui.adsabs.harvard.edu/abs/2021NatCo..12.1025Y}
}

@ARTICLE{Done2007,
       author = {{Done}, Chris and {Gierli{\'n}ski}, Marek and {Kubota}, Aya},
        title = "{Modelling the behaviour of accretion flows in X-ray binaries. Everything you always wanted to know about accretion but were afraid to ask}",
      journal = {\aapr},
         year = 2007,
        month = dec,
       volume = {15},
       number = {1},
        pages = {1-66},
          doi = {10.1007/s00159-007-0006-1},
archivePrefix = {arXiv},
       eprint = {0708.0148},
 primaryClass = {astro-ph},
       adsurl = {https://ui.adsabs.harvard.edu/abs/2007A&ARv..15....1D}
}

@ARTICLE{Parker2015,
       author = {{Parker}, M.~L. and {Tomsick}, J.~A. and {Miller}, J.~M. and {Yamaoka}, K. and {Lohfink}, A. and {Nowak}, M. and {Fabian}, A.~C. and {Alston}, W.~N. and {Boggs}, S.~E. and {Christensen}, F.~E. and {Craig}, W.~W. and {F{\"u}rst}, F. and {Gandhi}, P. and {Grefenstette}, B.~W. and {Grinberg}, V. and {Hailey}, C.~J. and {Harrison}, F.~A. and {Kara}, E. and {King}, A.~L. and {Stern}, D. and {Walton}, D.~J. and {Wilms}, J. and {Zhang}, W.~W.},
        title = "{NuSTAR and Suzaku Observations of the Hard State in Cygnus X-1: Locating the Inner Accretion Disk}",
      journal = {\apj},
         year = 2015,
        month = jul,
       volume = {808},
       number = {1},
          eid = {9},
        pages = {9},
          doi = {10.1088/0004-637X/808/1/9},
archivePrefix = {arXiv},
       eprint = {1506.00007},
 primaryClass = {astro-ph.HE},
       adsurl = {https://ui.adsabs.harvard.edu/abs/2015ApJ...808....9P}
}

@INPROCEEDINGS{Gendreau2016,
       author = {{Gendreau}, Keith C. and {Arzoumanian}, Zaven and {Adkins}, Phillip W. and {Albert}, Cheryl L. and {Anders}, John F. and {Aylward}, Andrew T. and {Baker}, Charles L. and {Balsamo}, Erin R. and {Bamford}, William A. and {Benegalrao}, Suyog S. and {Berry}, Daniel L. and {Bhalwani}, Shiraz and {Black}, J. Kevin and {Blaurock}, Carl and {Bronke}, Ginger M. and {Brown}, Gary L. and {Budinoff}, Jason G. and {Cantwell}, Jeffrey D. and {Cazeau}, Thoniel and {Chen}, Philip T. and {Clement}, Thomas G. and {Colangelo}, Andrew T. and {Coleman}, Jerry S. and {Coopersmith}, Jonathan D. and {Dehaven}, William E. and {Doty}, John P. and {Egan}, Mark D. and {Enoto}, Teruaki and {Fan}, Terry W. and {Ferro}, Deneen M. and {Foster}, Richard and {Galassi}, Nicholas M. and {Gallo}, Luis D. and {Green}, Chris M. and {Grosh}, Dave and {Ha}, Kong Q. and {Hasouneh}, Monther A. and {Heefner}, Kristofer B. and {Hestnes}, Phyllis and {Hoge}, Lisa J. and {Jacobs}, Tawanda M. and {J{\o}rgensen}, John L. and {Kaiser}, Michael A. and {Kellogg}, James W. and {Kenyon}, Steven J. and {Koenecke}, Richard G. and {Kozon}, Robert P. and {LaMarr}, Beverly and {Lambertson}, Mike D. and {Larson}, Anne M. and {Lentine}, Steven and {Lewis}, Jesse H. and {Lilly}, Michael G. and {Liu}, Kuochia Alice and {Malonis}, Andrew and {Manthripragada}, Sridhar S. and {Markwardt}, Craig B. and {Matonak}, Bryan D. and {Mcginnis}, Isaac E. and {Miller}, Roger L. and {Mitchell}, Alissa L. and {Mitchell}, Jason W. and {Mohammed}, Jelila S. and {Monroe}, Charles A. and {Montt de Garcia}, Kristina M. and {Mul{\'e}}, Peter D. and {Nagao}, Louis T. and {Ngo}, Son N. and {Norris}, Eric D. and {Norwood}, Dwight A. and {Novotka}, Joseph and {Okajima}, Takashi and {Olsen}, Lawrence G. and {Onyeachu}, Chimaobi O. and {Orosco}, Henry Y. and {Peterson}, Jacqualine R. and {Pevear}, Kristina N. and {Pham}, Karen K. and {Pollard}, Sue E. and {Pope}, John S. and {Powers}, Daniel F. and {Powers}, Charles E. and {Price}, Samuel R. and {Prigozhin}, Gregory Y. and {Ramirez}, Julian B. and {Reid}, Winston J. and {Remillard}, Ronald A. and {Rogstad}, Eric M. and {Rosecrans}, Glenn P. and {Rowe}, John N. and {Sager}, Jennifer A. and {Sanders}, Claude A. and {Savadkin}, Bruce and {Saylor}, Maxine R. and {Schaeffer}, Alexander F. and {Schweiss}, Nancy S. and {Semper}, Sean R. and {Serlemitsos}, Peter J. and {Shackelford}, Larry V. and {Soong}, Yang and {Struebel}, Jonathan and {Vezie}, Michael L. and {Villasenor}, Joel S. and {Winternitz}, Luke B. and {Wofford}, George I. and {Wright}, Michael R. and {Yang}, Mike Y. and {Yu}, Wayne H.},
        title = "{The Neutron star Interior Composition Explorer (NICER): design and development}",
    booktitle = {Space Telescopes and Instrumentation 2016: Ultraviolet to Gamma Ray},
         year = 2016,
       editor = {{den Herder}, Jan-Willem A. and {Takahashi}, Tadayuki and {Bautz}, Marshall},
       series = {Society of Photo-Optical Instrumentation Engineers (SPIE) Conference Series},
       volume = {9905},
        month = jul,
          eid = {99051H},
        pages = {99051H},
          doi = {10.1117/12.2231304},
       adsurl = {https://ui.adsabs.harvard.edu/abs/2016SPIE.9905E..1HG}
}

@ARTICLE{Jana2022,
       author = {{Jana}, Arghajit and {Naik}, Sachindra and {Jaisawal}, Gaurava K. and {Chhotaray}, Birendra and {Kumari}, Neeraj and {Gupta}, Shivangi},
        title = "{AstroSat observation of X-ray dips and state transition in the black hole candidate MAXI J1803-298}",
      journal = {\mnras},
         year = 2022,
        month = apr,
       volume = {511},
       number = {3},
        pages = {3922-3936},
          doi = {10.1093/mnras/stac315},
archivePrefix = {arXiv},
       eprint = {2202.00479},
 primaryClass = {astro-ph.HE},
       adsurl = {https://ui.adsabs.harvard.edu/abs/2022MNRAS.511.3922J}
}

@ARTICLE{Jana2020,
       author = {{Jana}, Arghajit and {Debnath}, Dipak and {Chatterjee}, Debjit and {Chatterjee}, Kaushik and {Chakrabarti}, Sandip Kumar and {Naik}, Sachindra and {Bhowmick}, Riya and {Kumari}, Neeraj},
        title = "{Accretion Flow Evolution of a New Black Hole Candidate MAXI J1348-630 during the 2019 Outburst}",
      journal = {\apj},
         year = 2020,
        month = jul,
       volume = {897},
       number = {1},
          eid = {3},
        pages = {3},
          doi = {10.3847/1538-4357/ab9696},
archivePrefix = {arXiv},
       eprint = {2004.03792},
 primaryClass = {astro-ph.HE},
       adsurl = {https://ui.adsabs.harvard.edu/abs/2020ApJ...897....3J}
}

@ARTICLE{Saha2023,
       author = {{Saha}, Debasish and {Mandal}, Manoj and {Pal}, Sabyasachi},
        title = "{Swift J1728.9-3613 is a black hole X-ray binary: a spectral and timing study using NICER}",
      journal = {\mnras},
         year = 2023,
        month = feb,
       volume = {519},
       number = {1},
        pages = {519-529},
          doi = {10.1093/mnras/stac3575},
archivePrefix = {arXiv},
       eprint = {2210.13748},
 primaryClass = {astro-ph.HE},
       adsurl = {https://ui.adsabs.harvard.edu/abs/2023MNRAS.519..519S}
}

@ARTICLE{Mandal2024,
       author = {{Mandal}, Manoj and {Saha}, Debasish and {Pal}, Sabyasachi and {Manna}, Arijit},
        title = "{Multi-wavelength observation of MAXI J1348─630 during the outburst in 2019}",
      journal = {\apss},
         year = 2024,
        month = feb,
       volume = {369},
       number = {2},
          eid = {18},
        pages = {18},
          doi = {10.1007/s10509-024-04280-z},
archivePrefix = {arXiv},
       eprint = {2104.09926},
 primaryClass = {astro-ph.HE},
       adsurl = {https://ui.adsabs.harvard.edu/abs/2024Ap&SS.369...18M}
}

@ARTICLE{Gierlinski2008,
       author = {{Gierli{\'n}ski}, Marek and {Done}, Chris and {Page}, Kim},
        title = "{X-ray irradiation in XTE J1817-330 and the inner radius of the truncated disc in the hard state}",
      journal = {\mnras},
         year = 2008,
        month = aug,
       volume = {388},
       number = {2},
        pages = {753-760},
          doi = {10.1111/j.1365-2966.2008.13431.x},
archivePrefix = {arXiv},
       eprint = {0803.0496},
 primaryClass = {astro-ph},
       adsurl = {https://ui.adsabs.harvard.edu/abs/2008MNRAS.388..753G}
}

@ARTICLE{Konig2026,
       author = {{K{\"o}nig}, Ole and {Steiner}, James F. and {Bollemeijer}, Niek and {Connors}, Riley M.~T. and {Dauser}, Thomas and {Dov{\v{c}}iak}, Michal and {Fan}, Ningyue and {Garc{\'\i}a}, Javier A. and {Horn}, David and {Ingram}, Adam and {Lucchini}, Matteo and {Mastroserio}, Guglielmo and {Miller}, Cal and {Nathan}, Edward and {Nowak}, Michael A. and {Pottschmidt}, Katja and {Remillard}, Ron and {Song}, Yujia and {Svoboda}, Ji{\v{r}}{\'\i} and {van der Klis}, Michiel and {Ubach}, Santiago and {Wilms}, J{\"o}rn and {Zhang}, Yuexin},
        title = "{Systematic Assessment of Disk Truncation in the Black Hole X-Ray Binary Swift J1727.8─1613 Using NICER}",
      journal = {\apj},
         year = 2026,
        month = may,
       volume = {1003},
       number = {1},
          eid = {73},
        pages = {73},
          doi = {10.3847/1538-4357/ae60f5},
archivePrefix = {arXiv},
       eprint = {2604.14693},
 primaryClass = {astro-ph.HE},
       adsurl = {https://ui.adsabs.harvard.edu/abs/2026ApJ..1003...73K}
}

@ARTICLE{Dunn2011,
       author = {{Dunn}, R.~J.~H. and {Fender}, R.~P. and {K{\"o}rding}, E.~G. and {Belloni}, T. and {Merloni}, A.},
        title = "{A global study of the behaviour of black hole X-ray binary discs}",
      journal = {\mnras},
         year = 2011,
        month = feb,
       volume = {411},
       number = {1},
        pages = {337-348},
          doi = {10.1111/j.1365-2966.2010.17687.x},
archivePrefix = {arXiv},
       eprint = {1009.2599},
 primaryClass = {astro-ph.HE},
       adsurl = {https://ui.adsabs.harvard.edu/abs/2011MNRAS.411..337D}
}

@ARTICLE{Tetarenko2019,
       author = {{Tetarenko}, A.~J. and {Casella}, P. and {Miller-Jones}, J.~C.~A. and {Sivakoff}, G.~R. and {Tetarenko}, B.~E. and {Maccarone}, T.~J. and {Gandhi}, P. and {Eikenberry}, S.},
        title = "{Radio frequency timing analysis of the compact jet in the black hole X-ray binary Cygnus X-1}",
      journal = {\mnras},
         year = 2019,
        month = apr,
       volume = {484},
       number = {3},
        pages = {2987-3003},
          doi = {10.1093/mnras/stz165},
archivePrefix = {arXiv},
       eprint = {1901.03751},
 primaryClass = {astro-ph.HE},
       adsurl = {https://ui.adsabs.harvard.edu/abs/2019MNRAS.484.2987T}
}

@ARTICLE{You2024,
       author = {{You}, Bei and {Yang}, Shuai-kang and {Yan}, Zhen and {Cao}, Xinwu and {Zdziarski}, Andrzej A.},
        title = "{The Delayed Radio Emission in the Black Hole X-Ray Binary MAXI J1348-630}",
      journal = {\apjl},
         year = 2024,
        month = jul,
       volume = {969},
       number = {2},
          eid = {L33},
        pages = {L33},
          doi = {10.3847/2041-8213/ad5b50},
archivePrefix = {arXiv},
       eprint = {2406.15994},
 primaryClass = {astro-ph.HE},
       adsurl = {https://ui.adsabs.harvard.edu/abs/2024ApJ...969L..33Y}
}

@INPROCEEDINGS{Shapopi2021,
       author = {{Shapopi}, J.~N.~S. and {Zdziarski}, A.},
        title = "{Correlations and lags between X-ray and radio emission of Cyg X-1}",
    booktitle = {High Energy Astrophysics in Southern Africa},
         year = 2021,
        month = jan,
          eid = {55},
        pages = {55},
          doi = {10.22323/1.371.0055},
archivePrefix = {arXiv},
       eprint = {2004.11786},
 primaryClass = {astro-ph.HE},
       adsurl = {https://ui.adsabs.harvard.edu/abs/2021heas.confE..55S}
}

@ARTICLE{Sridhar2020,
       author = {{Sridhar}, Navin and {Garc{\'\i}a}, Javier A. and {Steiner}, James F. and {Connors}, Riley M.~T. and {Grinberg}, Victoria and {Harrison}, Fiona A.},
        title = "{Evolution of the Accretion Disk-Corona during the Bright Hard-to-soft State Transition: A Reflection Spectroscopic Study with GX 339-4}",
      journal = {\apj},
         year = 2020,
        month = feb,
       volume = {890},
       number = {1},
          eid = {53},
        pages = {53},
          doi = {10.3847/1538-4357/ab64f5},
archivePrefix = {arXiv},
       eprint = {1912.11447},
 primaryClass = {astro-ph.HE},
       adsurl = {https://ui.adsabs.harvard.edu/abs/2020ApJ...890...53S}
}

@ARTICLE{Basak2017,
       author = {{Basak}, Rupal and {Zdziarski}, Andrzej A. and {Parker}, Michael and {Islam}, Nazma},
        title = "{Analysis of NuSTAR and Suzaku observations of Cyg X-1 in the hard state: evidence for a truncated disc geometry}",
      journal = {\mnras},
         year = 2017,
        month = dec,
       volume = {472},
       number = {4},
        pages = {4220-4232},
          doi = {10.1093/mnras/stx2283},
archivePrefix = {arXiv},
       eprint = {1705.06638},
 primaryClass = {astro-ph.HE},
       adsurl = {https://ui.adsabs.harvard.edu/abs/2017MNRAS.472.4220B}
}

@ARTICLE{Gehrels2004,
       author = {{Gehrels}, N. and {Chincarini}, G. and {Giommi}, P. and {Mason}, K.~O. and {Nousek}, J.~A. and {Wells}, A.~A. and {White}, N.~E. and {Barthelmy}, S.~D. and {Burrows}, D.~N. and {Cominsky}, L.~R. and {Hurley}, K.~C. and {Marshall}, F.~E. and {M{\'e}sz{\'a}ros}, P. and {Roming}, P.~W.~A. and {Angelini}, L. and {Barbier}, L.~M. and {Belloni}, T. and {Campana}, S. and {Caraveo}, P.~A. and {Chester}, M.~M. and {Citterio}, O. and {Cline}, T.~L. and {Cropper}, M.~S. and {Cummings}, J.~R. and {Dean}, A.~J. and {Feigelson}, E.~D. and {Fenimore}, E.~E. and {Frail}, D.~A. and {Fruchter}, A.~S. and {Garmire}, G.~P. and {Gendreau}, K. and {Ghisellini}, G. and {Greiner}, J. and {Hill}, J.~E. and {Hunsberger}, S.~D. and {Krimm}, H.~A. and {Kulkarni}, S.~R. and {Kumar}, P. and {Lebrun}, F. and {Lloyd-Ronning}, N.~M. and {Markwardt}, C.~B. and {Mattson}, B.~J. and {Mushotzky}, R.~F. and {Norris}, J.~P. and {Osborne}, J. and {Paczynski}, B. and {Palmer}, D.~M. and {Park}, H.-S. and {Parsons}, A.~M. and {Paul}, J. and {Rees}, M.~J. and {Reynolds}, C.~S. and {Rhoads}, J.~E. and {Sasseen}, T.~P. and {Schaefer}, B.~E. and {Short}, A.~T. and {Smale}, A.~P. and {Smith}, I.~A. and {Stella}, L. and {Tagliaferri}, G. and {Takahashi}, T. and {Tashiro}, M. and {Townsley}, L.~K. and {Tueller}, J. and {Turner}, M.~J.~L. and {Vietri}, M. and {Voges}, W. and {Ward}, M.~J. and {Willingale}, R. and {Zerbi}, F.~M. and {Zhang}, W.~W.},
        title = "{The Swift Gamma-Ray Burst Mission}",
      journal = {\apj},
         year = 2004,
        month = aug,
       volume = {611},
       number = {2},
        pages = {1005-1020},
          doi = {10.1086/422091},
archivePrefix = {arXiv},
       eprint = {astro-ph/0405233},
 primaryClass = {astro-ph},
       adsurl = {https://ui.adsabs.harvard.edu/abs/2004ApJ...611.1005G}
}

@ARTICLE{Zhang2023,
       author = {{Zhang}, Liang and {M{\'e}ndez}, Mariano and {Garc{\'\i}a}, Federico and {Zhang}, Yuexin and {Ma}, Ruican and {Altamirano}, Diego and {Yang}, Zi-Xu and {Ma}, Xiang and {Tao}, Lian and {Huang}, Yue and {Jia}, Shumei and {Zhang}, Shuang-Nan and {Qu}, Jinlu and {Song}, Liming and {Zhang}, Shu},
        title = "{Type-A quasi-periodic oscillation in the black hole transient MAXI J1348-630}",
      journal = {\mnras},
         year = 2023,
        month = dec,
       volume = {526},
       number = {3},
        pages = {3944-3950},
          doi = {10.1093/mnras/stad3062},
archivePrefix = {arXiv},
       eprint = {2310.04208},
 primaryClass = {astro-ph.HE},
       adsurl = {https://ui.adsabs.harvard.edu/abs/2023MNRAS.526.3944Z}
}

@ARTICLE{Miyamoto1993,
       author = {{Miyamoto}, Sigenori and {Iga}, Sayuri and {Kitamoto}, Shunji and {Kamado}, Yasuhide},
        title = "{Another Canonical Time Variation of X-Rays from Black Hole Candidates in the Very High Flare State?}",
      journal = {\apjl},
         year = 1993,
        month = jan,
       volume = {403},
        pages = {L39},
          doi = {10.1086/186716},
       adsurl = {https://ui.adsabs.harvard.edu/abs/1993ApJ...403L..39M}
}

@ARTICLE{Nespoli2003,
       author = {{Nespoli}, E. and {Belloni}, T. and {Homan}, J. and {Miller}, J.~M. and {Lewin}, W.~H.~G. and {M{\'e}ndez}, M. and {van der Klis}, M.},
        title = "{A transient variable 6 Hz QPO from GX 339-4}",
      journal = {\aap},
         year = 2003,
        month = dec,
       volume = {412},
        pages = {235-240},
          doi = {10.1051/0004-6361:20031423},
archivePrefix = {arXiv},
       eprint = {astro-ph/0309437},
 primaryClass = {astro-ph},
       adsurl = {https://ui.adsabs.harvard.edu/abs/2003A&A...412..235N}
}

@ARTICLE{Song2026,
       author = {{Song}, Yujia and {Steiner}, James F. and {Zhao}, Tong and {Zhang}, Yuexin and {Fan}, Ningyue and {K{\"o}nig}, Ole and {Ubach Ramirez}, Santiago and {Wong}, Josephine and {Gou}, Lijun and {Garcia}, Javier A.},
        title = "{Exploring the Spin Dependence on Mass Inclination and Distance for the Newly Discovered Black Hole X-Ray Binary Swift J151857.0─572147}",
      journal = {\apj},
         year = 2026,
        month = mar,
       volume = {999},
       number = {2},
          eid = {255},
        pages = {255},
          doi = {10.3847/1538-4357/ae4335},
archivePrefix = {arXiv},
       eprint = {2603.07908},
 primaryClass = {astro-ph.HE},
       adsurl = {https://ui.adsabs.harvard.edu/abs/2026ApJ...999..255S}
}

@ARTICLE{Wang2018,
       author = {{Wang-Ji}, Jingyi and {Garc{\'\i}a}, Javier A. and {Steiner}, James F. and {Tomsick}, John A. and {Harrison}, Fiona A. and {Bambi}, Cosimo and {Petrucci}, Pierre-Olivier and {Ferreira}, Jonathan and {Chakravorty}, Susmita and {Clavel}, Ma{\"\i}ca},
        title = "{The Evolution of GX 339-4 in the Low-hard State as Seen by NuSTAR and Swift}",
      journal = {\apj},
         year = 2018,
        month = mar,
       volume = {855},
       number = {1},
          eid = {61},
        pages = {61},
          doi = {10.3847/1538-4357/aaa974},
archivePrefix = {arXiv},
       eprint = {1712.02571},
 primaryClass = {astro-ph.HE},
       adsurl = {https://ui.adsabs.harvard.edu/abs/2018ApJ...855...61W}
}

\end{document}